\documentclass[journal,12pt,onecolumn,draftclsnofoot,]{IEEEtran}
\usepackage{lipsum}
\usepackage{amsfonts}
\usepackage{graphicx,tabularx}
\usepackage{epstopdf}
\usepackage{algorithmic}
\usepackage{array}
\usepackage{amsmath,graphicx,epsfig,float,subfigure,times,latexsym,verbatim,mathrsfs,amssymb,textcomp,txfonts,color}
\usepackage[flushleft]{threeparttable}
\usepackage{algorithm}
\usepackage{algorithmic}
\usepackage{url}
\usepackage{bbm}

\begin{document}
%
\title{Cooperative Raman Spectroscopy for Real-time In Vivo Nano-biosensing}


\author{Hongzhi Guo,~\IEEEmembership{Student Member,~IEEE,}
	Josep Miquel Jornet,~\IEEEmembership{Member,~IEEE,}\\
    Qiaoqiang Gan,~\IEEEmembership{Member,~IEEE,} and Zhi Sun,~\IEEEmembership{Member,~IEEE}
\thanks{This work was supported by the U.S. National Science Foundation (NSF) under Grant No. CBET-1445934.}
\thanks{The authors are with the Department of Electrical Engineering, University at Buffalo, the State University of New York, Buffalo, NY 14260, United States. E-mail:
\{hongzhig, jmjornet, qqgan, zhisun\}@buffalo.edu.}
}


\maketitle

\begin{abstract}
In the last few decades, the development of miniature biological sensors that can detect and measure different phenomena at the nanoscale has led to transformative disease diagnosis and treatment techniques. Among others, biofunctional Raman nanoparticles have been utilized in vitro and in vivo for multiplexed diagnosis and detection of different biological agents. However, existing solutions require the use of bulky lasers to excite the nanoparticles and similarly bulky and expensive spectrometers to measure the scattered Raman signals, which limit the practicality and applications of this nano-biosensing technique. In addition, due to the high path loss of the intra-body environment, the received signals are usually very weak, which hampers the accuracy of the measurements. In this paper, the concept of cooperative Raman spectrum reconstruction for real-time in vivo nano-biosensing is presented for the first time. The fundamental idea is to replace the single excitation and measurement points (i.e., the laser and the spectrometer, respectively) by a network of interconnected nano-devices that can simultaneously excite and measure nano-biosensing particles. More specifically, in the proposed system a large number of nanosensors jointly and distributively collect the Raman response of nano-biofunctional nanoparticles (NBPs) traveling through the blood vessels. This paper presents a detailed description of the sensing system and, more importantly, proves its feasibility, by utilizing accurate models of optical signal propagation in intra-body environment and low-complexity estimation algorithms. The numerical results show that with a certain density of NBPs, the reconstructed Raman spectrum can be recovered and utilized to accurately extract the targeting intra-body information.
\end{abstract}


\IEEEpeerreviewmaketitle
\section{Introduction}
Driven by the development of nanotechnology, emerging nanosensors have been envisioned to provide unprecedented sensing accuracy for many important applications, such as food safety detection \cite{nanoenvironment}, agriculture disease monitoring \cite{afsharinejad2016performance},
and health monitoring \cite{nguyen2015surface}, among others. Since nanosensors can interact directly with the most fundamental elements in matter, e.g., atoms and molecules, they can provide ultra-high sensitivity. One of the most promising applications of nanosensors is in vivo biosensing \cite{eckert2013novel,ramesh2013towards}, where nanosensors are injected into human body to collect real-time information. Nanosensors can be utilized both to detect well-known diseases at their very early stage as well as to provide new fundamental insights and understanding of biological processes that cannot be observed at the macroscopic level.

The use of nanoscale communication techniques can enable data transmission among nanosensors \cite{atakan2012body,akyildiz2012monaco}. Whether molecular, acoustic or electromagnetic, there are two fundamental limitations of directly using {\textit {active}} nanosensors in human body. First, wireless nanosensors require continuous power supply to support wireless data transmission and motion control. However, due to the limited size of the nanosensor, a large battery cannot be equipped and, even worse, recharging the battery is difficult. Second, the wireless nanosensor requires circuitry and antenna to process and radiate wireless signals, which further increases its size. In order to alleviate the side-effects caused by nanosensors in human body, we need to reduce its size by removing the battery and wireless components.

Metallic nanoparticles coated with Raman active reporter molecules have been widely used as surface enhanced Raman scattering labels for multiplexed diagnosis and bio-detection of DNA and proteins \cite{vijayarangamuthu2014nanoparticle,saha2012gold,henry2016surface}. This is a promising solution since it does not require power and wireless components on the nanoparticles. Their motion is driven by the dynamic fluids in human circular system and the information is delivered by electromagnetic scattering. The Raman active reporter molecules interact with chemicals inside human body and the incident single-frequency optical light is scattered into a wide frequency band with unique power spectrum due to molecule vibration. Based on this unique spectrum, we can identify the molecules. Although this approach suffers from low detected power due to the small scattering cross section, the scattering efficiency can be improved by placing the Raman active reporter molecules on the surface of metallic nanoparticles \cite{saha2012gold,kneipp1997single}.

While this solution can dramatically reduce the size of the nano-device that is injected into the human body, it still has limitations, which prohibit it from being widely used. First, a laser is needed to excite the engineered nanoparticle inside the human body and a spectrometer is demanded to detect scattered Raman signal. Both the laser and the spectrometer are bulky and expensive and, thus, are not portable or affordable. In addition, the accuracy of this sensing setup is not high enough since the scattered Raman signal is much weaker than the emitted signal by the laser due to the small scattering cross section of the nanoparticle and the dispersive and lossy propagation medium.

\begin{figure}[t]
  \centering
    \includegraphics[width=0.35\textwidth]{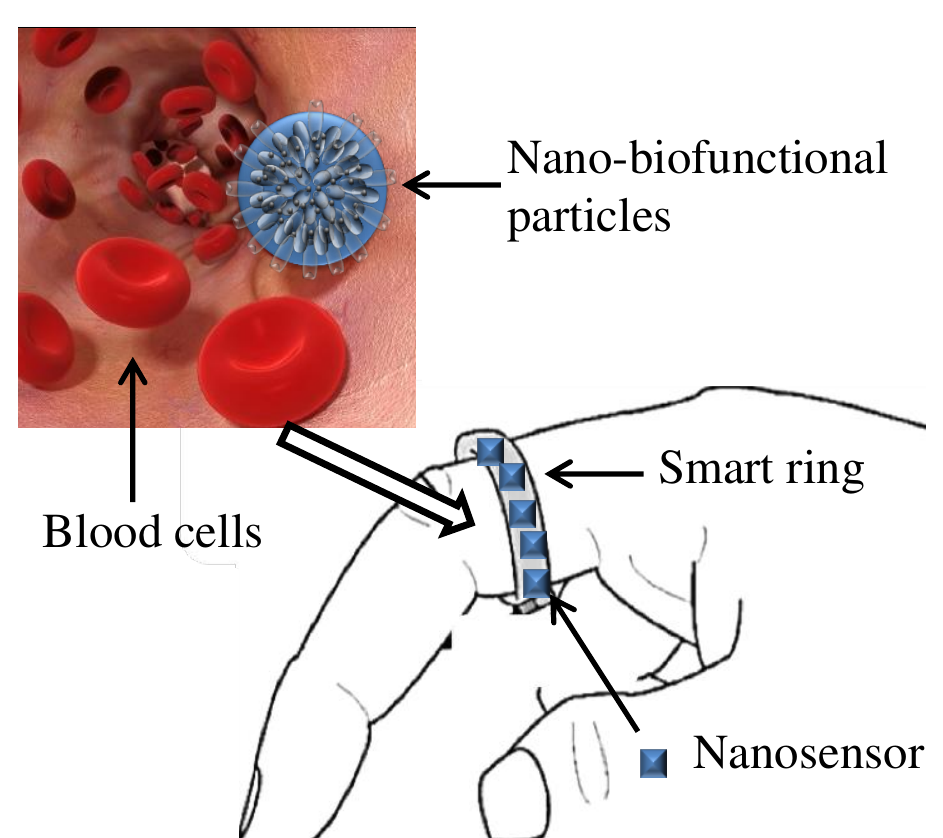}
    \vspace{-5pt}
  \caption{Cooperative Raman spectroscopy sensing system on a wearable smart ring. External part (the lower figure) is the nanosensors, which are installed on the smart ring; internal part (the upper figure) is the NBP flowing in blood vessels. }
  \vspace{-10pt}
  \label{fig:sys1}
\end{figure}

To address the aforementioned challenges, we propose the concept of cooperative Raman spectroscopy, which can be integrated on wearable devices \cite{pantelopoulos2010survey,rhee1998ring}, such as a smart nanophotonic ring. The system consists of external nanosensors and internal nano-biofunctional particles (NBPs), as shown in Fig.~\ref{fig:sys1}. The bulky expensive lasers and spectrometers are replaced with distributed nanosensors on a smart ring, which can both emit and detect optical signals, by leveraging the state of the art in nano-lasers and nano-photodetectors \cite{feng2014single,nafari2017modeling}. The nanosensors are placed on a smart ring which can reduce the distance to the intra-body particles to increase the received signal strength. Moreover, by installing nanosensors distributively, we can increase the diversity of detection and optimally allocate resources to make the sensing system more robust.

In this paper, we design a sensing system for cooperative Raman spectroscopy. More specifically, first, we present the system architecture and describe the processes of signal generation, scattering and detection. Based on the operational framework, we provide theoretical models to describe each part of the system, including signal propagation, noise, NBP density, and nanosensor's position. In addition, we provide detailed description of the information carried by NBPs and the method to extract the information. Different from conventional sensing systems, signals are not only distorted by the propagation channel, but also the molecular noise and shot noise are introduced. The limited power on the smart ring poses another challenge. Based on the system model, we derive the sensing capacity and define optimal power allocation schemes to increase the sensing accuracy in each sub-band of the Raman spectrum. Also, we derive the expected detected power of each nanosensor using the stochastic system model. Based on the theoretical model and nanosensor observations, we provide both centralized and distributed Raman spectrum estimation algorithm, from which the molecule information is extracted. The numerical simulation validates the accuracy of the proposed estimation methods.

The remaining part of this paper is organized as follows. The system architecture, operational framework, and system model are introduced in Section II. After that, the sensing capacity and optimal power allocation strategy are discussed in Section III. This is followed by the signal estimation algorithm presented in Section IV. The proposed system performance is numerically evaluated in Section V. Finally, this paper is concluded in Section VI.
\section{System Architecture and Model}
The system architecture of cooperative Raman spectroscopy consists of two important units, as shown in Fig.~\ref{fig:sys1}. The first key element is the external nanosensors on a smart ring, which are employed to 1) radiate optical signals, 2) detect scattered signals by NBPs, and 3) process the detected information to reconstruct the Raman spectrum. The second key component is the internal NBPs, which are injected into blood vessels to sense bioinformation. The bioinformation on NBPs can be extracted by using electromagnetic scattering. In the following, we first introduce the system architecture.

\subsection{System Architecture}
A NBP flowing in the human body can interact with different types of molecules. Once it is illuminated by a monochromatic (single frequency) optical signal, it absorbs the signal and scatters it into a wide spectrum. The spectrum is unique for different molecules due to their different chemical structures \cite{saha2012gold}. The objective of the proposed sensing system is to excite the NBP using a single-frequency optical signal and reconstruct the wide-band spectrum to identify the molecule. With this in mind, a large number of interconnected nanosensors are installed on a smart ring and each nanosensor has many nano-emitters and nano-detectors. In transmission, the nano-emitters generate and radiate the same monochromatic optical signal. In reception, due to the challenges in creating broadband detectors able to capture the entire Raman spectrum, each nano-detector is tuned to a different narrow sub-band and many of them are placed together on a nanosensor to cover the whole wide-band spectrum. The nanosensors are uniformly distributed on the ring. In this way, no matter how the ring is worn, it does not affect the sensing results.


\begin{figure}[t]
  \centering
  \subfigure[Centralized sensing architecture. All the nanosensors first send detected photon numbers to a data fusion center on the ring. Then, the data fusion center can either process or send the raw data to a smart phone. ]{
    \label{fig:sys2}
    \includegraphics[width=0.7\textwidth]{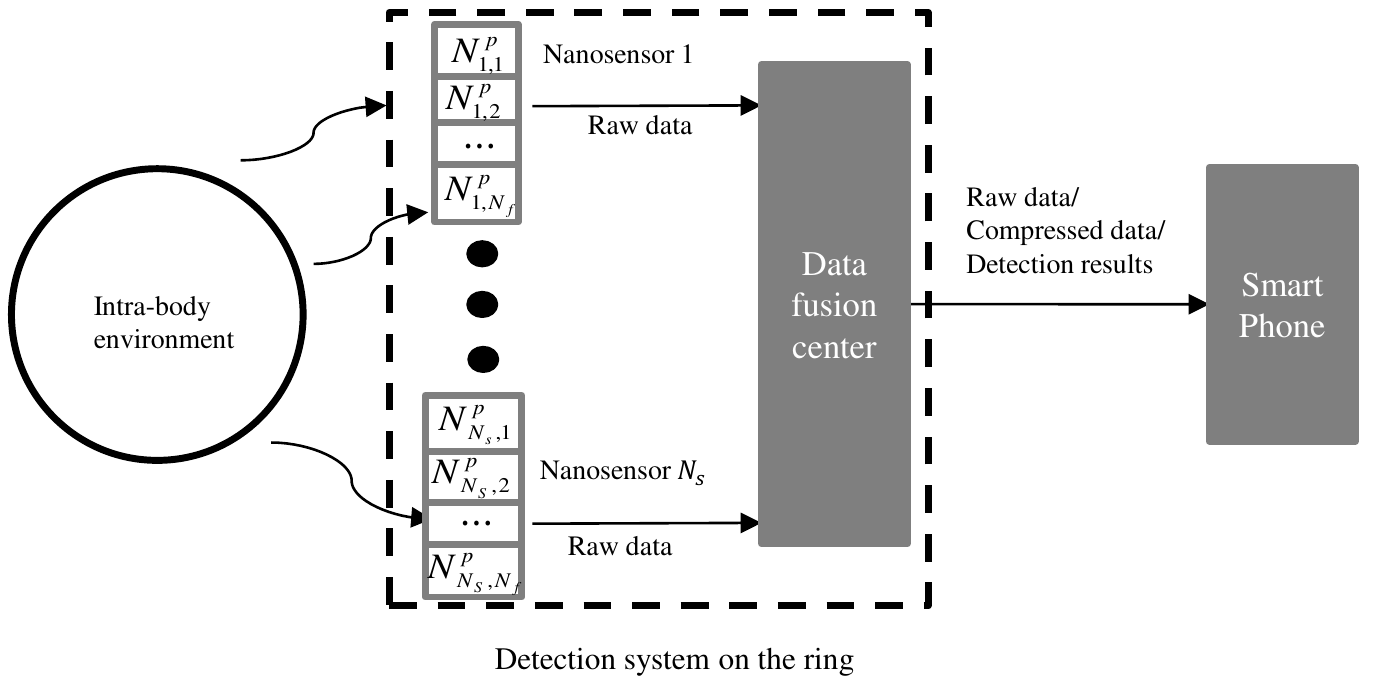}}\quad
  \subfigure[Distributed sensing architecture. Each nano-detector first processes the detected information and quantizes the estimated results locally. Then, the nanosensors send quantized results to a data fusion center on the smart ring to do a global estimation. Finally, the detection results are reported to a smart phone.]{
    \label{fig:sys3}
    \includegraphics[width=0.7\textwidth]{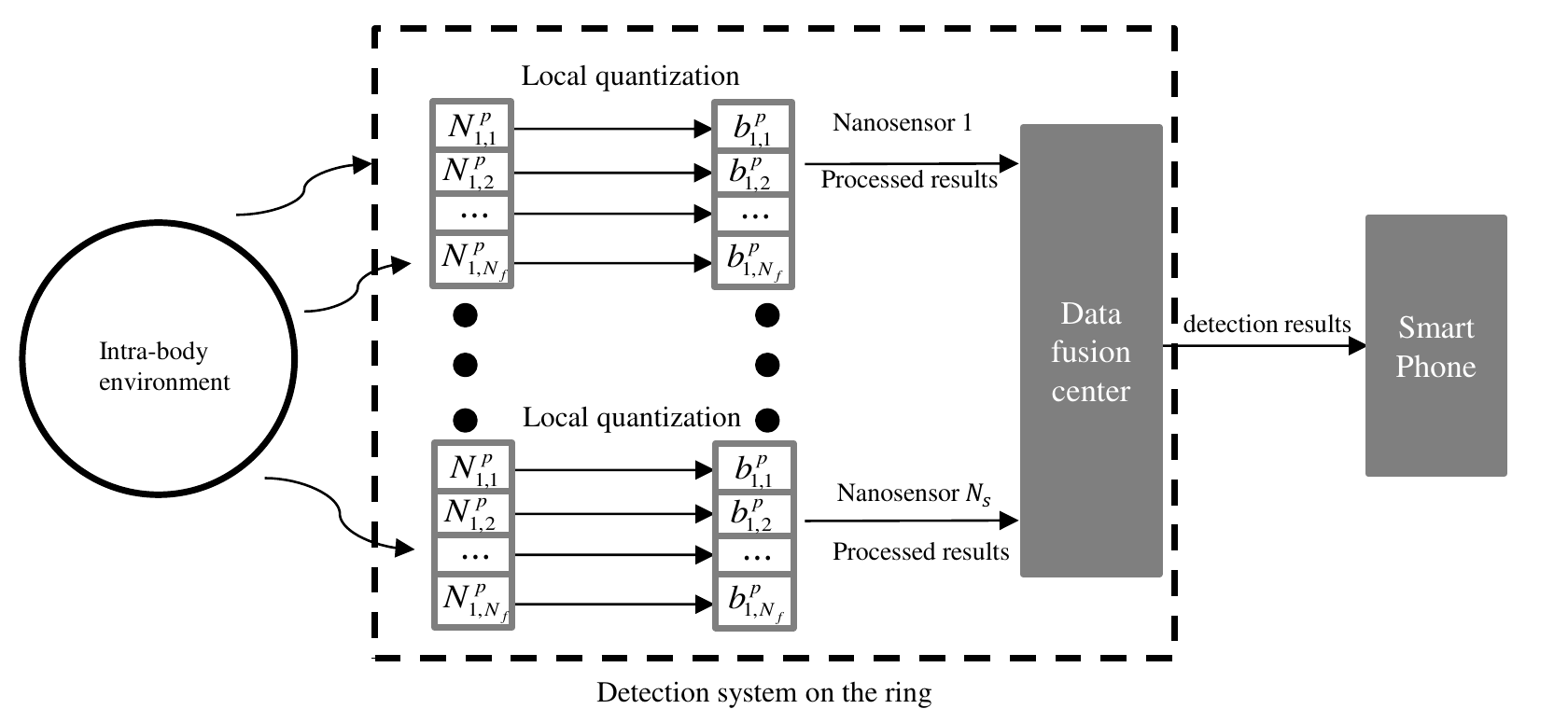}}\quad
      \vspace{-2pt}
  \caption{Sensing system architecture. $N_{i,j}^p$ is the detected photon number by the $i^{th}$ nanosensor's $j^{th}$ sub-band nano-detector. $b_{i,j}$ is the quantized estimated results by the $i^{th}$ nanosensor's $j^{th}$ sub-band nano-detector. }
    \vspace{-10pt}
  \label{fig:architecture}
\end{figure}

%

Once the raw spectrum data are collected by each nanosensor, there are primarily two approaches to reconstruct the spectrum and detect the molecules. 1) As shown in Fig.~\ref{fig:sys2}, the first one is a centralized architecture, where the raw data are sent directly to a data fusion center to do further processing and identification. This method can provide the most accurate results since all the raw data are considered in the estimation algorithm. Besides estimating the spectrum directly, the data fusion center can first compress the raw spectrum data and then send to the smart phone. In this way, the smart phone takes charge of spectrum reconstruction and molecule identification. However, there are two drawbacks which can prevent us from applying this architecture. First, the communication overhead is large since all the data need to be sent, which can increase the system delay and thus real-time detection may not be possible. The second drawback is that the signal processing in data fusion center requires a large amount of energy and computation resource which increases the burden of the ring. 2) The second architecture relies on a distributed sensing concept as shown in Fig.~\ref{fig:sys3}. Each of the nanosensor performs estimation algorithm and send the quantized single-bit results to the data fusion center. Based on the local results, the data fusion center performs a global estimation and identification and then send the results to the smart phone. In this way, most of the data are processed locally and thus the communication overhead can be dramatically reduced. Nevertheless, this system requires more computation resources for the nanosensor and the estimation accuracy may not be as high as the centralized system.

The operational framework of the cooperative Raman spectroscopy consists of three phases.
\begin{itemize}
\item  First, the synchronized nano-emitters on the smart ring radiate optical signals at the same frequency into the finger. The wavelength of the signal is usually between 450~nm to 1100~nm.
\item  Second, the flowing particles in blood vessels absorb the radiated optical signal from emitters. Then, the particles scatter the power into a wide spectrum.
\item  Lastly, the scattered signals propagate towards nano-detectors and then the nano-detectors operating at different frequencies receive the corresponding photons. After that, one can use different data fusion and sensing architectures as shown in Fig.~\ref{fig:sys2}, and Fig.~\ref{fig:sys3} to process the sensed data, upon which the Raman spectrum can be reconstructed and the machine learning algorithms can be applied to identify the category of the molecules.
\end{itemize}
Based on the sensing system architecture and operational framework, we provide the mathematical model for each component in the following.
\subsection{System Model}

\begin{figure}[t]
  \centering
    \includegraphics[width=0.65\textwidth]{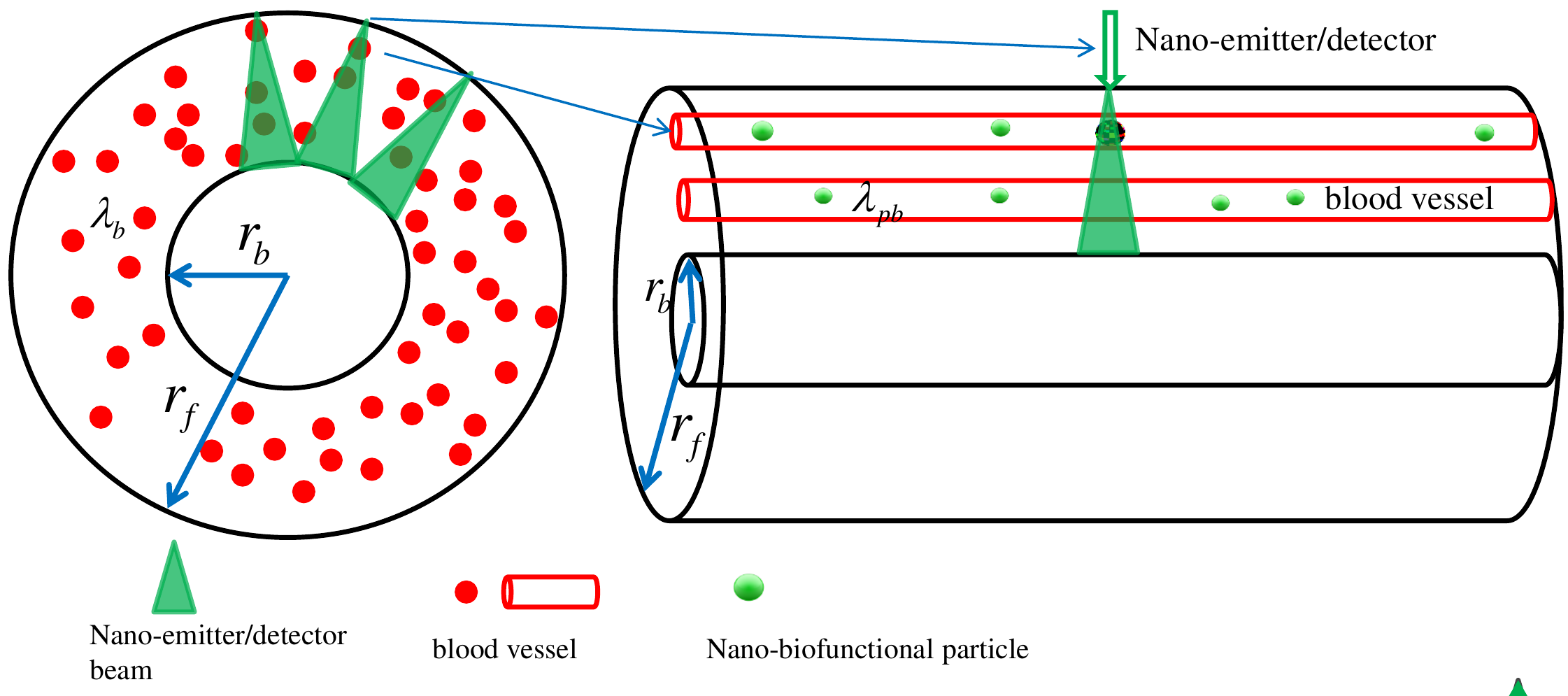}
    \vspace{-5pt}
  \caption{Illustration of light beam and detector's effective area. The left-hand side is vertical cross section of a finger. The outer circle is the cross section of finger and the inner one is the cross section of bone. The right-hand side is horizontal illustration of the finger and blood vessels. }
  \vspace{-10pt}
  \label{fig:cover}
\end{figure}

Consider that there are $N_{s}$ nanosensors uniformly installed on a ring and each nanosensor has $N_{f}$ pairs of nano-emitters and nano-detectors. The positions of a pair of nano-emitter and nano-detector are considered to be the same since they are very close to each other. The whole Raman spectrum is divided into $N_f$ sub-bands and each nano-detector on the nanosensor can detect signals in one sub-band. Note that due to the noise and low-density of NBPs, some detectors may not receive enough power and thus multiple nanosensors are employed to make the system reliable. Since the bone is relatively far from the skin and it is hard to penetrate, it can block the propagation of optical signal. We assume both the finger and the bone are cylinders with radius $r_f$ and $r_b$, respectively. The blood vessels, including artery, vein, and capillary, are randomly distributed between the skin and bone with density $\lambda_{b}$. In each blood vessel, the NBPs arrive with a density proportional to the area of the blood vessel's cross section, which is denoted by $\lambda_{pb}=\lambda_0 S_{b}$, where $\lambda_0$ is the NBP density of a unit area and $S_b$ is the area of a blood vessel's cross section. In reality, $\lambda_0$ is a function of time. When the NBPs are injected into the circulatory system, $\lambda_0$ gradually increases. After a while, some of the NBPs are disposed by natural physiological actions and the density gradually decreases. Due to the high directivity of the nano-emitter and nano-detector, we consider they can only radiate/detect signal with a large gain within a narrow beam. The system parameters are also depicted in Fig.~\ref{fig:cover} and the symbol notations are provided in Table~\ref{tab:notations}. In this paper, we consider the sensing is quasi-static since the optical light propagates much faster than NBPs' movement. Thus, in the following the NBPs are assumed to be static and the optical channel remains constant during the sensing period.

\begin{table}
\renewcommand{\arraystretch}{1.3}
\caption{Symbol Notations}
\label{tab:notations}
\centering
\begin{tabular}{c|c|c|c}
    \hline
    Symbols  &  Explanation&Symbols  &  Explanation\\
    \hline
    \hline

    $\lambda_0$  &   NBP density in a unit area & $S_b$& Area of a blood vessel's cross section\\
    \hline

    $\lambda_b$  &   Blood vessel density & $S_u$& Maximum area of a blood vessel's cross section\\
    \hline

   $\lambda_{pb}$  &   NBP density in a blood vessel &$S_l$ &Minimum area of a blood vessel's cross section\\
    \hline

    $r_f$  &   Radius of finger & $N_s$ & Nanosensor number\\
    \hline

    $r_b$  &   Radius of bone &$N_f$& Sub-band number\\
    \hline
    $d$  &   Distance to emitter/detector &$N_b$& Blood vessel number\\
    \hline
     $\eta_{f_t,f_j}$  &  {Scattering coefficient,  \newline input $f_t$, scattering $f_i$ }&$\alpha$& Emission/detection beam angle\\
    \hline
     $h_c$  &  Height of the beam &$l$& Length of a blood vessel covered by the beam\\
    \hline
     $\kappa$  &  Molecule noise &$\upsilon$& Dark current noise\\
    \hline
     $\sigma_c^2$  & Channel fading variance  &$\sigma_m^2$& Molecule noise variance\\
    \hline
\end{tabular}
\end{table}

\subsubsection{Signal Propagation Model}
The optical signals need to penetrate skin, fat, and blood vessels to reach the NBPs. Extensive analytical and empirical models have been derived to capture this process \cite{jacques2013optical,prahl1989monte,wang1995mcml,lin2011electromagnetic}. There are many categories of cells and tissues and their properties can be drastically different. In \cite{guo2016intra}, an analytical channel model for intra-body in vivo biosensing is developed by considering the properties of individual cells. In this paper, we use the same model to describe the propagation loss of EM wave radiated by the emitters, which can be simply written as
\begin{align}
\label{equ:channel}
h(f,d)=e^{\frac{-2}{(k_w r_c)^2}\sum_{n=1}^{N_{stop}}(2n+1)\Re\left({\mathcal F}_M^n+{\mathcal F}_N^n\right)d},
\end{align}
where $k_w$ is the propagation constant, $r_c$ is cell radius, $N_{stop}$ is the numerical calculation order, ${\mathcal F}_M^n$ and ${\mathcal F}_N^n$ are wave vector coefficients in \cite{guo2016intra}, and $\Re$ denotes the real part of a complex number, $d$ is the propagation distance and $f$ is the operating frequency. Besides this large scale fading, due to the multipath effect caused by scattering, a Rayleigh fading coefficient is also considered whose scale parameter is $\sigma_c$.

\subsubsection{Particle Scattering Coefficient and Quantization Model}

\begin{figure}[t]
  \centering
  \subfigure[Continuous Raman spectrum. ]{
    \label{fig:raman_shift}
    \includegraphics[width=0.3\textwidth]{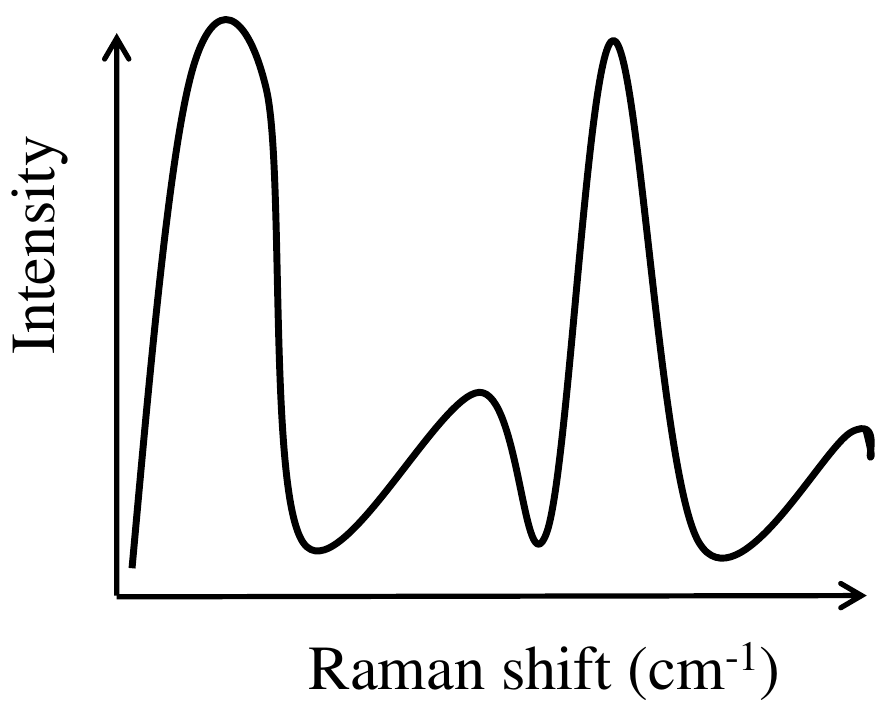}}\quad
  \subfigure[Discrete Raman spectrum.]{%
    \includegraphics[width=0.3\textwidth]{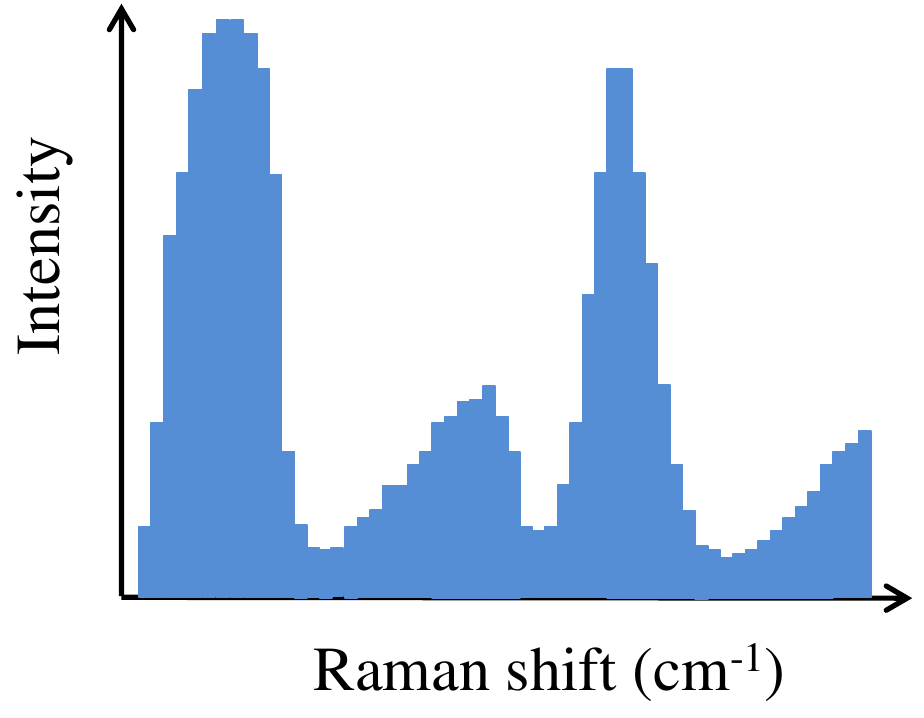}\quad
    \label{fig:raman_shift2}}
      \vspace{-5pt}
  \caption{Raman shift. }
    \vspace{-10pt}
  \label{fig:raman}
\end{figure}

The NBPs first absorb power from incident light and then scatter the power with unique information. Therefore, the NBPs can be regarded as an information source which sends encoded data $x$ to detectors. This process consists of two steps. First, the NBP absorb the incident signal power at frequency $f_t$. Then, the NBP reallocate the absorbed power based on the scattering coefficients $\eta_{f_t,f_j}$, where $f_j$ is the center frequency of a sub-band. Consequently, the scattered power forms a wide-band power spectrum that contains the information of the scattering coefficients.

As shown in Fig. \ref{fig:raman_shift}, the scattered signal by the NBP spreads on a wide spectrum with varied signal intensity. The intensity in the figure can be regarded as received power which is proportional to the particle scattering coefficient when the transmission power is given as a constant. This scattering coefficient is considered as the transmitted signal $x$. As shown in the figure, the spectrum is a continuous signal; however, the estimation is discrete, i.e., we can only estimate a single coefficient within a sub-band to approximate the continuously changed power spectrum, as depicted in Fig.~\ref{fig:raman_shift2}. As a result, we have to sample and quantize the continuous spectrum and then based on the scattering coefficient vector ${\boldsymbol \eta}=[\eta_{f_t,f_1}, \eta_{f_t,f_2},\cdots,\eta_{f_t,f_{N_f}}]$, we can reconstruct the Raman spectrum.

As mentioned in the system architecture, we can use both centralized and distributed system to estimate the coefficient. In the centralized algorithm, the number of bits that the nanosensor uses to describe the received signal can significantly affect the system communication overhead. In the distributed system, each nanosensor has their own estimation and the number of bits it utilizes is also crucial. To reduce the computation burden of nanosensors and the system communication overhead, we use simple binary quantization. When the scattering coefficient (Raman intensity) is higher than a threshold, $\eta_{f_t,f_j}$ is considered as 1. When the estimated coefficient is smaller than the threshold, the quantization process considers the scattering coefficient (Raman intensity) as 0. To estimate the value, we set several this kind of thresholds and divide the sensors into subgroups. Each subgroup has its own thresholds. Finally, based on the quantization results of all the nanosensors, we can estimate ${\boldsymbol \eta}$. The details will be discussed in the spectrum estimation in Section~\ref{sec:estimation}. In addition, since different molecules have different spectrum, the event of transmitting 1 or 0 is a random process. In the following, we consider the probability of transmitting 1 is $p$ and the probability of transmitting 0 is $1-p$.

\subsubsection{Noise Model}
The noise in a sensing system can corrupt the detected signals and significantly affect the sensing capability. In the cooperative Raman spectroscopy system, there are primarily two noises, namely, molecules noise and shot noise.

The NBPs flow through the circulatory system and interact with plenty of molecules. On one hand, they meet with the valuable molecules carrying health information. Through optical scattering, we can detect those molecules by identifying the power spectrum. On the other hand, the NBPs also encounter many unexpected molecules in intra-body environment. Although the particles are not designed to interact with these molecules, some chemical reactions can happen and change the particles' properties randomly, which are reflected in the received power spectrum. The original power spectrum is corrupted by unexpected noise power. Therefore, this noise needs to be taken into account when reconstruct the power spectrum.

Since the molecules in human body have a large variety of categories which demonstrate different resonant frequencies in Raman spectrum, we can consider the noise power is the same for all the frequency bands. Therefore, the noise can be considered as white with uniform power across a wide band. Due to the large amount of molecules, the noise value can be positive or negative, i.e., enhance or cancel the original resonance due to chemical reactions, and its distribution is Gaussian with mean value 0 and standard deviation $\sigma_m$. Consequently, the noise caused by molecules can be regarded as additive white Gaussian noise $\kappa^m \sim \mathcal{N}(0, {\sigma_m}^2)$. With this molecules noise, the scattering coefficient of the biofunctional particle can be written as $\eta_{f_t,f_j}+\kappa^m=\eta_{f_t,f_j}(1+\kappa)$. Note that if $1+\kappa<0$ we consider the total scattering coefficient as 0.

Shot noise is dominant in the detector which obeys Poisson distribution. Let $x(t)=\eta_{f_t,f_j}(1+\kappa(t))$ be the scattered coefficient of a NBP plus molecule noise, $P^t$ be the emitter transmission power, and $h(f,d,t)$ be the response of the channel from nano-emitter to particle and then from particle to nano-detector. The received signal at a nano-detector by using direct detection can be written as
\begin{align}
\label{equ:sig}
y(t)=h(f,d,t)x(t)P^t+\upsilon(t),
\end{align}
where $\upsilon(t)$ is the dark current. Then, the light strength can be converted into doubly-stochastic Poisson process, which represents the number of photons arriving at the detector in a time interval $\Delta t$. The probability that there are $N_p$ photons arrive within $\Delta t$ is \cite{21285}
\begin{align}
\label{equ:prob_photon_num}
Pr \{{\hat y}(t+\Delta t)-{\hat y}(t)=N_p\}=\frac{e^{-\gamma_p}\cdot \gamma_p^{N_p}}{N_p!},
\end{align}
where ${\hat y}(t)$ is the converted $y(t)$ from light strength to photon intensity and
\begin{align}
\gamma_p=\int_{t}^{t+\Delta t}\left[y(t)\right] dt \approx \Delta t y(t),
\end{align}
where the approximation can be applied when $\Delta t$ is small enough. Note that here $\upsilon$ is a nonnegative constant \cite{shamai1993bounds} and $y$ can be taken to have units photons per second at the operating wavelength \cite{haas2003capacity}.

\subsubsection{Particle Arriving Model}

\begin{figure}[t]
  \centering
    \includegraphics[width=0.13\textwidth]{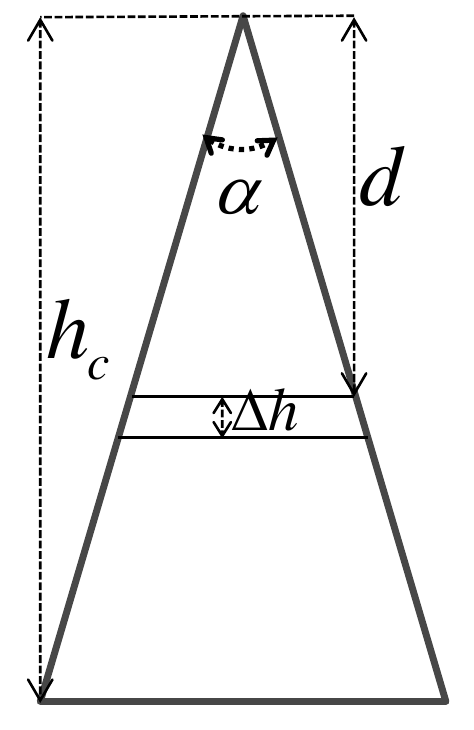}
    \vspace{-5pt}
  \caption{Geometry of nano-emitter/detector beam.}
  \vspace{-10pt}
  \label{fig:beam}
\end{figure}
The NBPs are injected into circulatory system with a certain density. They arrive at the targeting sensing area with a diluted density. To model this process, we consider the arrival rate of NBPs in a unit cross section of blood vessel is $\lambda_0$. Since different blood vessels have different cross section areas, their NBPs arrival rates are also different. Moreover, the process of NBP moving is modeled as Poisson process since the NBPs are independent and random in the blood \cite{erdmann1978recurrence}. The number of the NBPs that can be excited by the nano-emitter depends on the position of the blood vessel, the distance to the nano-emitter, and the density of NBPs. The radiated optical signal by a nano-emitter can cover a three dimensional cone and each nano-detector can receive the scattered optical signal in the same cone since the nano-emitter and nano-detector have almost the same position. As shown in Fig.~\ref{fig:cover}, the blood vessels are homogeneously distributed between skin and bone. Although NBPs can receive power from multiple beams as long as the nano-emitters are close enough to each other, we consider adjacent nano-emitters with overlapped beams work in different time slots to eliminate the correlation among them to reduce the complexity of analysis, i.e., in each time slot the NBPs within a beam can only receive power from one nano-emitter. Since the beam angle is small, we safely assume that all the NBPs on the same horizontal plane of the cone have the same distance to the emitter. For instance, the NBPs within $\Delta h$ in Fig.~\ref{fig:beam} have the same distance to the nano-emitter. To find the number of particles in a blood vessel and the received power, we need to find the distributions of the length of blood vessels within a cone and their distance to the nano-emitter. Given the blood vessel's effective length $l$ and its cross section area, the number of particles within it is given as
\begin{align}
Pr(n_p=N_p|L=l, S_b=s_b)=\frac{(\lambda_0s_bl /u)^{N_p}}{{N_p}!}e^{-\lambda_0s_b l/u},
\end{align}
where $s_b$ is the cross section area of the blood vessel and $u$ is the velocity of blood. We assume the cross section of the blood vessel is uniformly distributed in $[S_l, S_u]$ with a probability density function $f(s_b)=1/(S_u-S_l)$.
\subsubsection{Nanosensor Position and Minimum Number}
\begin{figure}[t]
  \centering
    \includegraphics[width=0.75\textwidth]{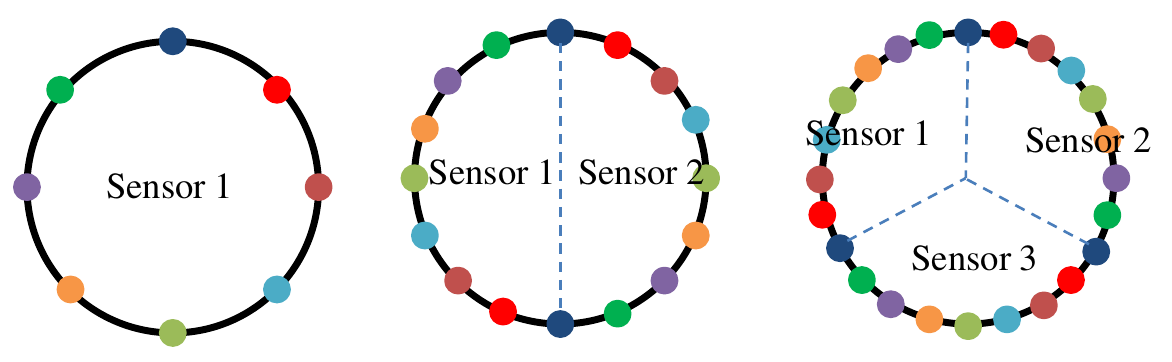}
    \vspace{-5pt}
  \caption{Distribution of nanosensors on the ring. $N_s=$ 1, 2, and 3 (it can be even more). Different colors in the figure denote different sub-bands. Only 8 sub-bands are depicted in the figure as an example.}
  \vspace{-10pt}
  \label{fig:deployment}
\end{figure}
One of the design objectives is that no matter how the smart ring is worn, it does not affect its performance. With this mind, we place the nanosensors in a homogeneous way as shown in Fig.~\ref{fig:deployment}. When there are $N_f$ sub-bands and $N_s$ nanosensors, we first place the nanosensors in sub-band 1 at $[0, \frac{2\pi}{N_s}, \cdots, \frac{2(N_s-1)\pi}{N_s}]$. Then, the nanosensors in sub-band 2 are placed at $[\frac{2\pi}{N_s N_f}, \frac{2\pi}{N_s}+\frac{2\pi}{N_s N_f},\cdots,\frac{2(N_s-1)\pi}{N_s}+\frac{2\pi}{N_s N_f}]$. Similarly, the nanosensors in the $n^{th}$ sub-band are placed at $[\frac{2(n-1)\pi}{N_s N_f}, \frac{2\pi}{N_s}+\frac{2(n-1)\pi}{N_s N_f},\cdots,\frac{2(N_s-1)\pi}{N_s}+\frac{2(n-1)\pi}{N_s N_f}]$. Three examples are provided in Fig.~\ref{fig:deployment} when $N_f=8$ and $N_s=1$, $2$, and $3$, respectively.

As discussed in preceding sections, within each nano-emitter/detector's beam, it is possible that there is no blood vessel. As a result, the nano-detector cannot receive any signal. If this happens for all the nano-detectors in a sub-band, the power spectrum of that sub-band is missing. Herein, in our design we need to guarantee that this can only happen with arbitrarily low probability. The blood vessels are homogeneously distributed and thus the probability that there are $N_b$ blood vessels within the effective area of $f_j$ sub-band is
\begin{align}
Pr(n_b=N_b)=\frac{(0.5\lambda_b N_s h_c^2 \tan\frac{\alpha}{2})^{N_b}}{N_b!}e^{-0.5\lambda_b N_s h_c^2 \tan\frac{\alpha}{2}}.
\end{align}
Thus, when $N_b=0$,
\begin{align}
Pr(n_b=0)=e^{-0.5\lambda_b N_s h_c^2 \tan\frac{\alpha}{2}}.
\end{align}
Since the blood vessel density is a constant number, which we cannot freely adjust, and the nano-emitter/detector's beamwidth is preconfigured, only the number of nanosensor can be varied. An arbitrarily small threshold $\tau_b$ is set to guarantee that $Pr(n_b=0)\leq\tau_b$ and the minimum sensor number is
\begin{align}
\label{equ:sensor_number}
N_b^s \geq \frac{-2 \ln \tau_b}{\lambda_b h_c^2 \tan\frac{\alpha}{2}}.
\end{align}
Note that this minimum number can only promise that there are blood vessels going through a nano-emitter's/detector's beam. It does not guarantee that the detector can receive scattered signal, because this also depends on NBP's density.

\section{Nanosensor Optimal Power Allocation}
Similar as other wearable devices, the power consumption is also a critical issue for the smart ring utilized for cooperative Raman spectroscopy \cite{martin2000issues}. In this section, we first derive a capacity for optical signal transmission in intra-body environment to measure the information delivered by a sub-band, upon which we develop the optimal power allocation scheme. In this paper, both the power and photon intensity are utilized. As described in \eqref{equ:sig}, the received signal can be expressed by the input signal and the dark current, which are both denoted in photon intensity. The photon intensity can be converted into power by multiplying the energy per photon $E_p=h_{PC}c_{LT}/\lambda_w$, where $h_{PC}$ is Planck's constant, $c_{LT}$ is the speed of light, and $\lambda_w$ is the wavelength.

\subsection{Capacity Analysis}
The capacity analysis is mainly based on \eqref{equ:sig}. Since the detection takes very short time, we assume the particle movement and channel status within such a period is constant and thus the time $t$ is neglected. When the nano-detector receives one photon, it considers the scattering coefficient as 1, which can be related to the results after quantization. Otherwise, the nano-detector considers the scattering coefficient as 0. Following the method in \cite{wyner1988capacity}, when the nano-detector receives more than 1 photon, the signal is regarded as 0 by considering it as an error. Since we consider a very shot period, the probability of receiving more than one photon is extremely low. If 0 is transmitted, we can only receive 0, which delivers no information. Consequently, we consider the scenario when 1 is received the transition probability of a sub-band channel is
\begin{align}
Pr(1|0)&=(h_{i,j,k} \cdot \kappa^m_{i,j,k} \cdot P^t_{i,j}+\upsilon )\cdot \delta_t \cdot e^{-(h_{i,j,k} \cdot \kappa^m_{i,j,k} \cdot P^t_{i,j}+\upsilon )\cdot \delta_t };\\
Pr(1|1)&=\left[h_{i,j,k}  \cdot(\eta_{f_t,f_j}+\kappa^m_{i,j,k})\cdot P^t_{i,j}+\upsilon\right] \cdot \delta_t \cdot e^{-\left[h_{i,j,k}  \cdot(\eta_{f_t,f_j} +\kappa^m_{i,j,k})\cdot P^t_{i,j}+\upsilon\right] \cdot \delta_t},
\end{align}
where $i$ is from 1 to $N_s$ (nanosensor number), $j$ is from 1 to $N_f$ (sub-band number), $k$ is from 1 to $N_p^{i,j}$ (NBP number in a nano-emitter's/detector's beam), and $h_{i,j,k}=h(f,d^{ep})\cdot h(f,d^{pd})$, where $d^{ep}$ and $d^{pd}$ are distance from nano-emitter to NBP and distance from NBP to nano-detector, respectively. Then, the mutual information can be written as
\begin{align}
I(X,Y)&=H\{Y\}-H\{Y|X\}\nonumber\\
&=H\left\{p \cdot Pr(1|1)+(1-p) \cdot Pr(1|0)\right\}-p \cdot H\left\{Pr(1|1)\right\}-(1-p) \cdot H\left\{Pr(1|0)\right\}.
\end{align}

As pointed out in \cite{wyner1988capacity}, $\delta_t$ is very small and two approximations can be made to simplify $I(X,Y)$, i.e., $H\{x\}=-x\log x+x$ and $e^{x\delta_t}\approx 1$. In addition, we define the following three functions:
\begin{align}
&\xi_1(x_1,x_2,x_3)=-(x_1 +x_2 +x_3)\log(x_1 +x_2 +x_3);\\
&\xi_2(x_1,x_2,x_3,x_4)=x_1(x_2  +x_3 +x_4) \log(x_2  +x_3 +x_4);\\
&\xi_3(x_1,x_2)=(1-x_1)x_2\log(x_2).
\end{align}
As a result, the ergodic capacity of the information within $\delta_t$ that we can obtain from the Raman signal is
\begin{align}
C=\max_{x(t)\leq \eta_{f_t,f_j}} E\{ \frac{I(X,Y)}{\delta_t}\}&\approx E\left\{\xi_1(ph_{i,j,k}P^t_{i,j} \eta_{f_t,f_j}, h_{i,j,k}P^t_{i,j}\kappa^m_{i,j,k},\upsilon )\nonumber \right.\\
&\left.+\xi_2(p,h_{i,j,k}P^t_{i,j} \eta_{f_t,f_j},h_{i,j,k}P^t_{i,j}\kappa^m_{i,j,k},\upsilon)+\xi_3(p,h_{i,j,k} \kappa^m_{i,j,k} P^t_{i,j}+\upsilon )\right\}.
\end{align}

Up to this point, we implicitly assume $N_p^{i,j}=1$, i.e., there is only one NBP within the nano-emitter/detector's beam cone. When there are multiple NBPs, the transition probability can be updated as
\begin{align}
Pr(1|0)&=(\sum_{k=1}^{N_p^{i,j}}h_{i,j,k} \cdot \kappa^m_{i,j,k} \cdot P^t_{i,j}+\upsilon )\cdot \delta_t \cdot e^{-(\sum_{k=1}^{N_p^{i,j}}h_{i,j,k} \cdot \kappa^m_{i,j,k} \cdot P^t_{i,j}+\upsilon )\cdot \delta_t };\\
Pr(1|1)&=\left[\sum_{k=1}^{N_p^{i,j}}h_{i,j,k}  \cdot(\eta_{f_t,f_j}+\kappa^m_{i,j,k})\cdot P^t_{i,j}+\upsilon\right] \cdot \delta_t \cdot e^{-\left[\sum_{k=1}^{N_p^{i,j}}h_{i,j,k}  \cdot(\eta_{f_t,f_j} +\kappa^m_{i,j,k})\cdot P^t_{i,j}+\upsilon\right] \cdot \delta_t}.
\end{align}

When there are $N_s$ nanosensors and each nanosensor has $N_f$ sub-bands, the system ergodic capacity can be written as

\begin{align}
\label{equ:sys_capacity}
C_{sys}=\sum_{i=1}^{N_s}\sum_{j=1}^{N_f}C_{i,j}\approx&\sum_{i=1}^{N_s}\sum_{j=1}^{N_f}E\left\{\xi_1(\sum_{k=1}^{N_p^{i,j}}ph_{i,j,k}P^t_{i,j} \eta_{f_t,f_j}, \sum_{k=1}^{N_p^{i,j}}h_{i,j,k}P^t_{i,j}\kappa^m_{i,j,k},\upsilon )\nonumber \right.\\
&\left.+\xi_2(p,\sum_{k=1}^{N_p^{i,j}}h_{i,j,k}P^t_{i,j} \eta_{f_t,f_j},\sum_{k=1}^{N_p^{i,j}}h_{i,j,k}P^t_{i,j}\kappa^m_{i,j,k},\upsilon)+\xi_3(p,\sum_{k=1}^{N_p^{i,j}}h_{i,j,k} \kappa^m_{i,j,k} P^t_{i,j}+\upsilon )\right\}.
\end{align}
Based on this equation, in the next section, we try to optimally allocate $P_{i,j}^t$ to achieve the best estimation results.

\subsection{Optimal Power Allocation}
\label{sec:opt_power}
Since the Raman spectrum occupies a wide frequency band and different frequencies experiences different absorption and scattering, it is inefficient to allocate the same amount of power to all the nano-emitters. In addition, according to the capacity analysis, if we allocate the same amount of power to each sub-band, the detected information volume are different, which leads to different accuracies. In other words, some of the sub-bands are highly distorted (i.e., the results are not trustable), but other sub-bands have well detected results. As a result, the whole reconstructed spectrum is not homogeneous in accuracy. When the emission power is large enough, there is no need to consider this problem since all the sub-bands have good enough accuracy. However, for the proposed cooperative Raman spectroscopy, the smart ring has very limited power and thus the emission power need to be as small as possible. In the following we derive an optimal power allocation scheme based on the developed capacity to efficiently utilize the power. Due to the unique sensing system, we do not have real-time channel state information and thus the power allocation is based on prior knowledge of the channel which is derived in \cite{guo2016intra} and experimental measurement in \cite{jacques2013optical}. Let the total sensing power in the ring be $P^t$. Since the nanosensors have the same sub-band emitters and detectors, the power can be first equally allocated to each nanosensor and then optimally allocated to each nano-emitter. Therefore, the transmission power of each nanosensor is $P^s=P^t/N_s$ and we can optimize the power allocation in one nanosensor instead of all the nanosensors, i.e., $C_{sys}\approx N_s\sum_{j=1}^{N_f}C_{j}$. We implicitly assume all the nanosensors have the same configuration and the subscript $i$ is neglected.

To guarantee that all the sub-bands have the same capability to extract information from the biofunctional particle, their capacity should be the same. Thus, the condition need to be satisfied is
\begin{align}
\label{equ:capacity_opt}C_{1}&=C_{2}=\cdots=C_{{N_f}},\\
\label{equ:opt_cons1}s.t.~&\sum_{j=1}^{N_f}P_j^t=P^s,
\end{align}
where $P_j^t$ is the $j^{th}$ sub-band emitter transmission power. By observing \eqref{equ:sys_capacity}, we can find that the transmission power is integrated with the channel condition. If all the $h_{j,k}P^t_{j}$ can be the same, then \eqref{equ:capacity_opt} can be satisfied. As a result, the optimal power for the $j^{th}$ sub-band can be given as
\begin{align}
\label{equ:opt_power}
P_j^t=\frac{P_s}{N_f \sum_{k=1}^{N_p^{j}} h_{j,k}}.
\end{align}
Since the $N_p^j$ and $h_{j,k}$ are dynamic random variables, which are determined by the NBPs. As discussed before, the power allocation is based on prior knowledge of the channel. Therefore, by using the system model provided in Section II, we derive the expected value of $\sum_{k=1}^{N_p^{j}} h_{j,k}$, which can eliminate the randomness in power allocation.

When the transmission power of an emitter is $P_j^t$, the detected power without noise can be written as
\begin{align}
\label{equ:detected}
P_j^d=\sum_{k=1}^{N_p^{j}} h_{j,k} \eta_{f_t,f_j} P_j^t.
\end{align}
In view of \eqref{equ:detected}, if we can find $P_j^d$ given $\eta_{f_t,f_j}$ and $P_j^t$, $\sum_{k=1}^{N_p^{j}} h_{j,k}$ can be found. It is worth noting that, since the bandwidth $B_{sub}=f_{i+1}-f_i$ is small enough, the channel can be considered as flat fading within a sub-band. Also, our analysis is general, which holds for all the sub-bands. We first derive the expected detected power for one nano-detector. As shown in Fig.~\ref{fig:beam}, we divide the cross section of the cone into sub-regions with height $\Delta h$. Then, we classify the NBPs into each sub-region based on their position. Here, the height $\Delta h$ is considered as the largest height of the blood vessel's cross section which is $\Delta h=2\sqrt{\frac{S_u}{\pi}}$. The expected detected power can be expressed as
\begin{align}
E\{P_j^d\}&=E\{\sum_{k=1}^{N_p^{j}}P_{j,k}^d\}\label{equ:average_rp}\\
&\approx E\{\sum_{n=1}^{R_s}\sum_{k=1}^{{\hat N}_{p_n}^j} P_{j,n,k}^d\}\approx \sum_{n=1}^{R_s} E\{{\hat N}_{p_n}^j\} E\{{\hat{P}_{j,n}^d}\},\label{equ:region_app}
\end{align}
where $P_{j,k}^d$ is the detected power scattered by the $k^{th}$ NBP, ${\hat N}_{p_n}$ is the NBP number within the $n^{th}$ sub-region, and ${\hat{P}_{j,n}^d}$ is the expected detected power scattered by the $n^{th}$ sub-region. Due to the division of the cross section of the beam cone, \eqref{equ:average_rp} can be approximated by \eqref{equ:region_app}. Next, we look at each sub-region and find the expected detected power.


In each sub-region, we consider all the NBPs have the same distance to the nano-detector since the beam angle is very small. The expected NBP number in a sub-region can be found by using
\begin{align}
E\{{\hat N}_{p_n}^j\}&=\sum_{n=1}^{\infty}\left[ n \cdot Pr({\hat N}_{p_n}^j=n)\right].
\end{align}
Due to the complicated blood vessel distribution and their different cross section areas, here we consider an equivalent scenario, i.e., the randomly distributed blood vessels in the same sub-region of the cone are considered as one equivalent blood vessel. The average length of a blood vessel in a sub-region can be expressed as
\begin{align}
{\hat l}&=\int_{0}^{d \tan{\frac{\alpha}{2}}}\frac{2\sqrt{(d \tan{\frac{\alpha}{2}})^2-x^2}}{d\tan{\frac{\alpha}{2}}}dx=\frac{\pi d \tan{\frac{\alpha}{2}}}{2}.
\end{align}
The cross section of the equivalent blood vessel can be approximated by $\frac{S_u+S_l}{2}$ since the cross section is uniformly distributed. The expected number of blood vessels in a sub-region can be expressed as
\begin{align}
\lambda_{eq}=\frac{\lambda_b d \Delta h \tan{\frac{\alpha}{2}}}{2 r_f^2}.
\end{align}
Then, the length of the equivalent blood vessel is $l_{eq}={\hat l}\cdot \lambda_{eq}$ and the probability that there are $n$ NBPs in the equivalent blood vessel can be written as
\begin{align}
\label{equ:number_exp}
Pr({\hat N}_{p_n}^j=n)=\frac{(\lambda_0s_{eq}l_{eq} /u)^n}{n!}e^{-\lambda_0s_{eq} l_{eq}/u}.
\end{align}
Next, the expected detected power from one particle at distance $d$ is given as
\begin{align}
\label{equ:power_exp}
E\{{\hat{P}_{j,n}^d}\}=E\{P_j^t G_t(f_t) h_{j,n} \eta_(f_t,f_j) G_r(f_j)\},
\end{align}
where $G_t(f_t)$ is the gain of the nano-emitter at frequency $f_t$ and $G_r(f_j)$ is the gain of the nano-detector. Since on the left-hand side of \eqref{equ:power_exp} only $h_{j,n}$ is a random variable (it is a function of distance and subject to Rayleigh fading), \eqref{equ:power_exp} can be simplified as
\begin{align}
\label{equ:power_exp_2}
E\{{\hat{P}_{j,n}^d}\}=\frac{\pi}{2}P_j^t  G_t(f_t) \eta_{f_t,f_j} G_r(f_j) h_{j,n} \sigma^2.
\end{align}
By substituting \eqref{equ:number_exp} and \eqref{equ:power_exp_2} into \eqref{equ:region_app}, we can obtain the expected detected power by a nano-detector. Finally, the expected value of $\sum_{k=1}^{N_p^{j}} h_{j,k}$ can be found by dividing the $E\{P_j^d\}$ by $\eta_{f_t,f_j} P_j^t$.

Different from conventional wireless communication using water-filling algorithm to optimally allocate power \cite{tse2005fundamentals}, the power allocation in \eqref{equ:opt_power} is inversely proportional to the channel condition. Often in wireless communications more power is given to the sub-bands with less attenuation to increase the system output. In this sensing system if more power is given to the sub-bands with less attenuation we can obtain accurate estimation results. However, those high attenuation sub-bands with less allocated power may generate unexpected peaks which makes it hard to identify the molecules. For instance, we express the idea by using a simplified notation ${\tilde{P}}_t {\tilde{h}} {\tilde{\eta}} + {\tilde{n}}={\tilde{P}}_r$ where ${\tilde{P}}_t$, ${\tilde{h}}$, ${\tilde{\eta}}$, ${\tilde{n}}$, and ${\tilde{P}}_r$ are transmission power, channel coefficient, scattering coefficient, system noise, and received power, respectively. First, if we use water-filling algorithm, when ${\tilde{h}}$ is large, ${\tilde{P}}_t $ is also large and thus ${\tilde{n}}$ is relatively small when compared with ${\tilde{P}}_t $. Hence, ${\tilde{\eta}}$ can be accurately estimated by using maximum likelihood ${\tilde{P}}_r/({\tilde{P}}_t  {\tilde{h}})$. When ${\tilde{h}}$ is small, ${\tilde{P}}_t $ is also small according to water-filling algorithm. The estimation becomes not accurate, especially when the noise is strong (i.e., received power is large) the estimated ${\tilde{\eta}}$ deviates a lot from the original value which generates a peak/null in the spectrum. Since identifying Raman spectrum mainly based on the resonant peaks, these unexpected peaks can cause misleading detection results. Consequently, the conventional water-filling algorithm does not work here and we need to allocate power following \eqref{equ:opt_power}.

The above power allocation does not include the scattering coefficient and we only use the channel condition due to the following reasons. Since the variation of the scattering coefficient is much larger than the distortion of the channel, the power allocation strategy is mainly affected by the scattering coefficients. In other words, the variation of $\eta_{f_t,f_j}$ is larger than $h(f,d)$ and thus the emitter transmission power is almost inversely proportional to the scattering coefficient. When the noise is small or transmission power is high enough, the estimation accuracy can be reasonable. However, when the system becomes highly distorted, the detected signal can be considered as noise. When we calculate the scattering coefficient, the transmission power need to be divided. Then we have two scenarios. First, when the detected signal variation is smaller than the scattering coefficient, this yields the original spectrum which is mainly the scattering coefficient. Thus, if we want to detect a molecule and allocate power based on its scattering coefficient, no matter what kind of molecules are inside human body, the detected results is always positive. Second, when the detected signal variation is large, since the noise is strong the scattering coefficient cannot be recovered. Generally, the sensing system fails at high noise. Since sometimes we can obtain positive detection results when noise is strong, in power allocation we do not consider the scattering coefficient and only the channel dispersion is taken into account. In addition, the optimal power allocation strategy is not affected by the quantization threshold; it is only determined by the optical channel condition.

\section{Spectrum Reconstruction}
\label{sec:estimation}
In this section, we provide both the centralized and distributed sensing algorithms to reconstruct the Raman spectrum based on the observations of nanosensors. Within the sensing period, the photon number received by nano-detectors is ${\bf N}^d\in \mathbb{R}^{N_s \times N_f}$ and its element $(i,j)$ means the received photon number by the $i^{th}$ sensor's $j^{th}$ nano-detector. Based on it, we estimate the NBP scattering coefficient ${\boldsymbol \eta}$ to find the Raman intensity.

\subsection{Spectrum Estimation with Shot Noise}
The detected photon is a random number according to \eqref{equ:prob_photon_num}. Based on the photon number, we need to estimate the received signal $y$ in \eqref{equ:sig}. As suggested by \eqref{equ:prob_photon_num}, the relation between the received signal and the photon number obeys Poisson distribution. Then, maximum likelihood can be utilized to estimate the received signal. We define
\begin{align}
g=\frac{e^{-y}\cdot y^{N^d_{i,j}}}{N^d_{i,j}!}\approx\frac{e^{-y}}{\sqrt{2\pi}}(ye)^{N^d_{i,j}} ({N^d_{i,j}})^{-N^d_{i,j}-\frac{1}{2}}.
\end{align}
Note that we consider the time interval $\Delta t$ is a constant and $\gamma_p$ is simply approximated by $y$. Then, we can obtain the derivative with respect to $N^d_{i,j}$,
\begin{align}
\label{equ:likelihood}
(\ln g)'=\ln(ye)+\frac{1}{2N^d_{i,j}}-\ln N^d_{i,j}-1.
\end{align}
The estimated received signal ${\hat y}$ which can maximize \eqref{equ:likelihood} is
\begin{align}
\label{equ:estimate_shot}
{\hat y}=e^{\ln N^d_{i,j}-\frac{1}{2 N^d_{i,j}}}\approx N^d_{i,j}.
\end{align}
The estimation mean square error can be written as
\begin{align}
e_a=\sum_{y=0}^{\infty}\left( \frac{e^{-y}\cdot y^{N^d_{i,j}}}{N^d_{i,j}!}\cdot( y-N^d_{i,j} )^2 \right).
\end{align}
Once we have the estimation of ${\hat y}$, we need to estimate the coefficient $\eta_{f_t,f_j}$ based on the knowledge of the system model. The shot noise $\upsilon$ is a a nonnegative constant \cite{shamai1993bounds} which can be subtracted from ${\hat y}$ and the channel information can be found by using the derived expected detected power, upon which we can estimate $\eta_{f_t,f_j}$.

\subsection{Scattering Coefficient Estimation}
\subsubsection{Centralized Sensing}
Up to this point, we have the knowledge of the received signal, shot noise, and expected value of the received power. Then, an estimation of $\eta_{f_t,f_j}$ can be written as
\begin{align}
\label{equ:estimate}
{\hat \eta}_{f_t, f_j}=\frac{\sum_{i=1}^{N_s}({\hat y}_{i,j}-\upsilon)^{+}}{ E\{\sum_{k=1}^{N_p^{j}} h_{j,k}\}\cdot {\hat N_s}}=\eta_{f_t,f_j}+\Delta n,
\end{align}
where $\Delta n$ is the estimation error, ${\hat y}_{i,j}$ is the estimated signal of the $i^{th}$ nanosensor's $j^{th}$ sub-band nano-detector, $(x)^+=max(0,x)$, ${\hat N_s}$ is the number that ${\hat y}_{i,j}-\upsilon\geq0$, and $E\{\sum_{k=1}^{N_p^{j}}h_{j,k}\}$ can be found via \eqref{equ:average_rp} to \eqref{equ:power_exp_2}.

In the centralized architecture, each nano-detector sends the received photon number to the data fusion center directly. Based on ${\bf N}^d$ and each detector's operating frequency, the received signal ${\hat y}_{i,j}$ can be first estimated using \eqref{equ:estimate_shot}. Then, the signal denoted by photon number is converted to power. The data fusion center can directly use \eqref{equ:estimate} to estimate the scattering coefficient. The centralized sensing algorithm is summarized in Algorithm~\ref{alg:centralized}. As we can see, the centralized sensing is very simple and it relies on the full information of all the sensed data which results in high communication overhead and high power consumption.

\begin{algorithm}
 \caption{Centralized Sensing}
 \begin{algorithmic}[1]
 \label{alg:centralized}
 \renewcommand{\algorithmicrequire}{\textbf{Input:}}
 \renewcommand{\algorithmicensure}{\textbf{Output: }}
 \REQUIRE ${\bf N}^d$, $E\{\sum_{k=1}^{N_p^{j}} h_{j,k}\}$, $N_s$, $\upsilon$
 \ENSURE  ${\hat {\boldsymbol \eta}}$
  \STATE Based on ${\bf N}^d$ estimate received signal ${\hat y}_{i,j}$
  \STATE Using \eqref{equ:estimate} to find global estimated ${\hat {\boldsymbol \eta}}$
 \end{algorithmic}
 \end{algorithm}

\subsubsection{Distributed Sensing}
Different from the centralized esitmation, in distributed estimation each nanosensor's detector first estimate and quantize the scattering coefficient. Only one bit is sent to the data fusion center for final spectrum reconstruction. In this way, the data communication overhead among nanosensors and data fusion center can be significantly reduced. Although we do not have the knowledge of the PDF (probability density function) of $\Delta n$, we can still estimate $\eta_{f_t,f_j}$ by using the method in \cite{ribeiro2006bandwidth}. However, different from \cite{ribeiro2006bandwidth}, the scattering coefficient is in $[0, \infty)$, i.e., it cannot be negative. Therefore, the algorithm need to be updated to apply it in Raman spectrum reconstruction. It should be noted that we assume the sensors have prior knowledge of the coefficient $\eta_{f_t,f_j}$, i.e., the sensing system tries to detect whether a molecule is in intra-body environment or not.

From the perspective of a nano-detector, it has the information of the detector's shot noise $\upsilon$, detected photon number $N^d_{i,j}$, the expected channel condition $E\{\sum_{k=1}^{N_p^{j}} h_{j,k}\}$, and the corresponding targeting NBP's $\eta_{f_t,f_j}$, where $f_j$ is its detecting center frequency. First, by using the detected photon number and \eqref{equ:estimate_shot}, the nano-detector can find the received signal and convert it into power notation ${\hat y}_{i,j}$. Then, it can estimate $\eta_{f_t,f_j}$ locally by using
\begin{align}
\label{equ:local}
{\hat \eta}_{f_t,f_j}^{local}=\frac{{\hat y}_{i,j}-\upsilon}{E\{\sum_{k=1}^{N_p^{j}} h_{j,k}\}}
\end{align}
Now, instead of sending ${\hat \eta}_{f_t,f_j}^{local}$ to the data fusion center, the nano-detector first quantize it and the quantization threshold is determined by the nanosensor.

The $N_s$ nanosensors are divided into $K$ groups and the group $G_k$ uses $\tau_k$ as quantization threshold. Each $\tau_k$ is considered as a threshold for binary quantization. Consider that the nanosensor collects the local estimation results and set the maximum quantization threshold as
\begin{align}
T_i=\max(\eta_{f_t,f_j})+\sum_{j=1}^{N_f}{\hat \eta}_{f_t,f_j}^{local}/N_f.
\end{align}
Ideally, $\max(\eta_{f_t,f_j})$ is the maximum value of the coefficient. However, due to the noise, dynamic NBP number, and channel distortion, the estimated value may be larger or smaller than the original scattering coefficient and different nanosensors may have drastically different estimated values, although the reconstructed spectrum may have similar shape. Then, the mean estimated scattering coefficient is added to adjust the level of the threshold. As a result, $Pr({\hat \eta}_{f_t,f_j}>T)\approx 0$. The interval $[0,T]$ is divided into $K$ sub-intervals $[\tau_{i,0},\tau_{i,1},\cdots, \tau_{i,K}]$, where $\tau_{i,K}=T_i$. Then, the nano-detector can quantize ${\hat \eta}_{f_t,f_j}^{local}$ using the thresholds.

The estimation of ${\hat \eta}_{f_t, f_i}$ can be updated as
\begin{align}
{\hat \eta}_{f_t,f_j}=\frac{1}{4}\sum_{k=1}^{K}\left\{\frac{1}{N_{G_k}}\sum_{s=1}^{N_{G_k}}\left[b_{s,j}(\tau_{i,k+1}-\tau_{i,k-1})\right]\right\},
\end{align}
where $N_{G_k}$ is the number of nanosensors in group $k$, whose estimated received signal $\hat{y}$ is not zero. The distributed sensing algorithm is summarized in Algorithm~\ref{alg:distributed}. In Algorithm~\ref{alg:distributed}, the steps from 1 to 11 are performed by the nano-detector and the step 12 is conducted in the data fusion center.

\begin{algorithm}
 \caption{Distributed Sensing}
 \begin{algorithmic}[1]
 \label{alg:distributed}
 \renewcommand{\algorithmicrequire}{\textbf{Input:}}
 \renewcommand{\algorithmicensure}{\textbf{Output: }}
 \REQUIRE ${\boldsymbol \eta}$,${\bf N}^d$, ${\bf h}_i^d$, $N_s$ nanosensors divided into $K$ groups and each group is $G_k$, $k=1,\cdots, K$
 \ENSURE  ${\hat {\boldsymbol \eta}}$
  \STATE Based on ${\bf N}^d$ estimate received signal $y_{i,j}$
  \STATE Using \eqref{equ:local} to find local estimated ${\hat \eta}_{f_t,f_j}^{local}$
  \STATE $T_{i}=\max(\eta_{f_t,f_j}+\sum_{j=1}^{N_f}{\hat \eta}_{f_t,f_j}^{local}/N_f)$
  \STATE $\tau_{i,k}=\frac{(k-1)T_{i}}{K},~k=1,\cdots, K$
  \FOR {nanosensors in $G_k$}
  \IF {${\hat \eta}_{f_t,f_j}^{local}<\tau_{i,k}$}
  \STATE $b_{i,j}=0$
  \ELSE \STATE $b_{i,j}=1$
  \ENDIF
  \ENDFOR
  \STATE ${\hat \eta}_{f_t,f_j}=\frac{1}{4}\sum_{k=1}^{K}\left\{\frac{1}{N_{G_k}}\sum_{s=1}^{N_{G_k}}\left[b_{s,j}(\tau_{i,k+1}-\tau_{i,k-1})\right]\right\}$
 \end{algorithmic}
 \end{algorithm}

\subsubsection{Estimation Error Evaluation}
By using the preceding estimated scattering coefficients of the NBP we can find the Raman intensity in the $j^{th}$ sub-band by using $I_R=\frac{P^t_{exp}\eta_{f_t,f_j}\lambda_j}{h_{PC} c_{LT}}$, where $P^t_{exp}$ is the transmission power used by the experiment in \cite{zhang2015ultrabroadband}, and $\lambda_j$ is the wavelength of the $j^{th}$ sub-band.

Note that, identifying the molecule is mainly based on the resonant peaks in the Raman spectrum and thus the level of the intensity is not crucial (i.e., it can also be adjusted by using different transmission power). Motivated by this observation, we first normalize the spectrum by dividing its mean value, then calculate the Mean Square Error (MSE), i.e.,
\begin{align}
e_s=\frac{1}{N_f}\sum_{j=1}^{N_f} \left(\frac{{I}_{R,j}}{{\bar I}_{R}}-\frac{{\hat I}_{R,j}}{{\bar {\hat I}}_R}\right)^2,
\end{align}
where $I_{R,j}$ is the original Raman intensity in the $j^{th}$ sub-band, ${\bar I}_R$ is the mean value of the original Raman intensity across all the sub-bands, ${\hat I}_{R,j}$ is the estimated Raman intensity in the $j^{th}$ sub-band, ${\bar {\hat I}}_R$ is the mean value of the estimated Raman intensity across all the sub-bands. The outage probability is defined as $Pr(e_s>\tau_t)$, where $\tau_t$ is a threshold. When $e_s$ is smaller than $\tau_t$, we consider the estimated results can maintain a certain accuracy. In the numerical analysis of the system performance and optimal configuration, we use the outage probability as a guideline.
\section{Numerical Analysis and Optimal System Configuration}
In this section, we try to find the optimal configuration of the system based on the system model and developed estimation algorithm. The optimal configuration design is constrained by the total amount of transmission power $P^t$ and the maximum number of nano-emitter and nano-detector. The optimal configuration of the system should meet three objectives, namely, 1) minimum number of nanosensors to ensure that we can successfully reconstruct the spectrum; 2) minimum NBP density to guarantee the accuracy and reliability of the estimation results; 3) minimum transmission power to reduce the overall power consumption of the system.

Before embarking on the analyses of different system configurations, we give an ideal estimated spectrum which has the optimized numbers of nanosensors, NBP density, and transmission power. Also, the considered molecule noise and shot noise power are relatively small. In this way, we show the characteristics of good estimations and then in the following discussions we investigate the effect of each parameter and find out their optimal values.
\subsection{Ideal Estimation}
\begin{figure}[t]
  \centering
    \includegraphics[width=0.6\textwidth]{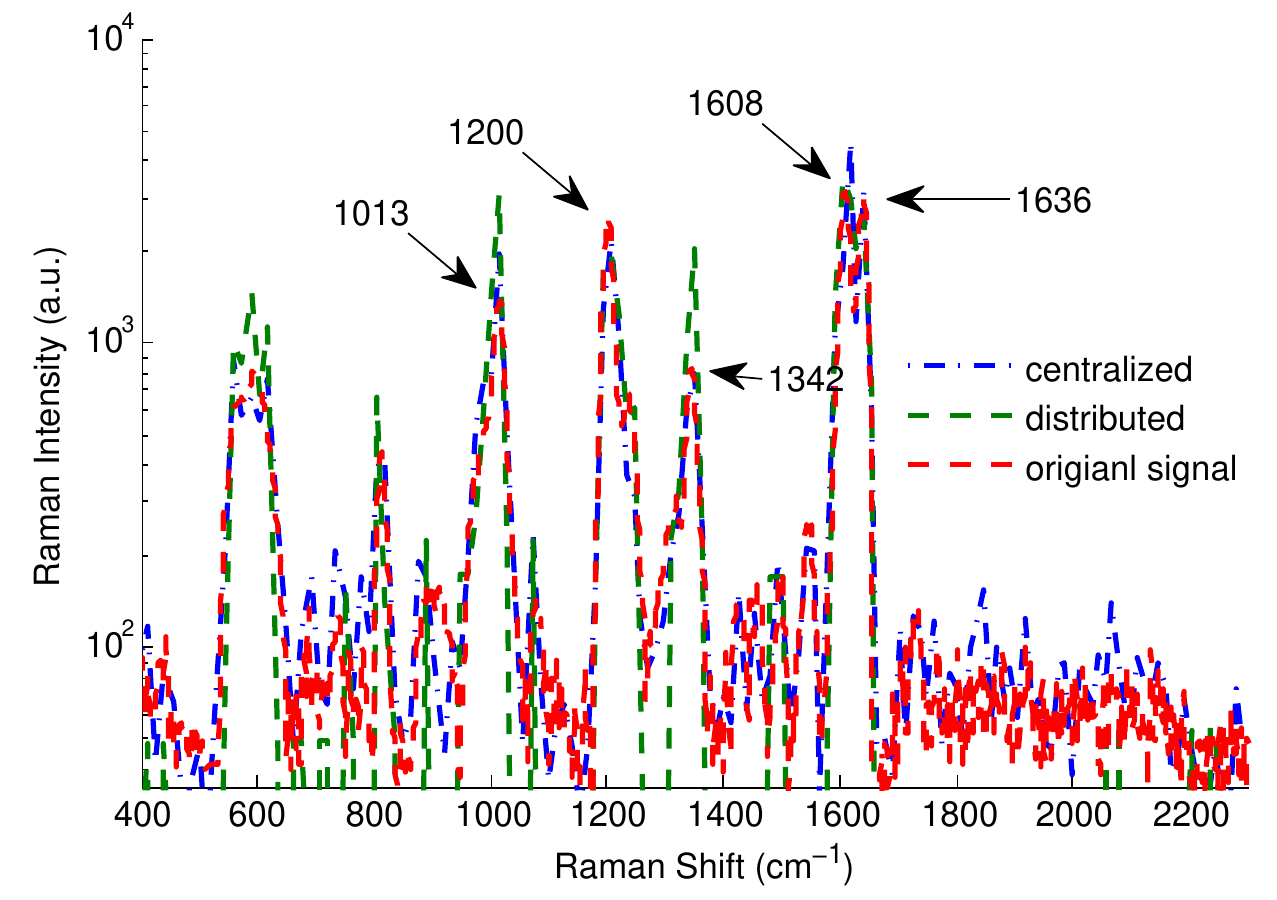}
    \vspace{-5pt}
  \caption{Estimated Raman intensity. The original signal is measured Raman intensity of 1,2-bits(4-pyridyl)-ethylene molecules in \cite{zhang2015ultrabroadband}. The intensity is displayed in log scale.}
  \vspace{-10pt}
  \label{fig:ideal_large_bw}
\end{figure}
The molecule utilized in this numerical simulation is 1,2-bits(4-pyridyl)-ethylene and its scattering coefficient and Raman spectrum are measured in \cite{zhang2015ultrabroadband}. In the numerical analysis, we first randomly generate a set of blood vessels but we do not change their positon and number in the following numerical analysis since the blood vessels are fixed in reality. Other random parameters such as NBP density and position, channel fading, and noise are randomly generated in each numerical simulation. The numerical parameters are provided in Table~\ref{tab:parameter}.

\begin{figure}[t]
  \centering
    \includegraphics[width=0.8\textwidth]{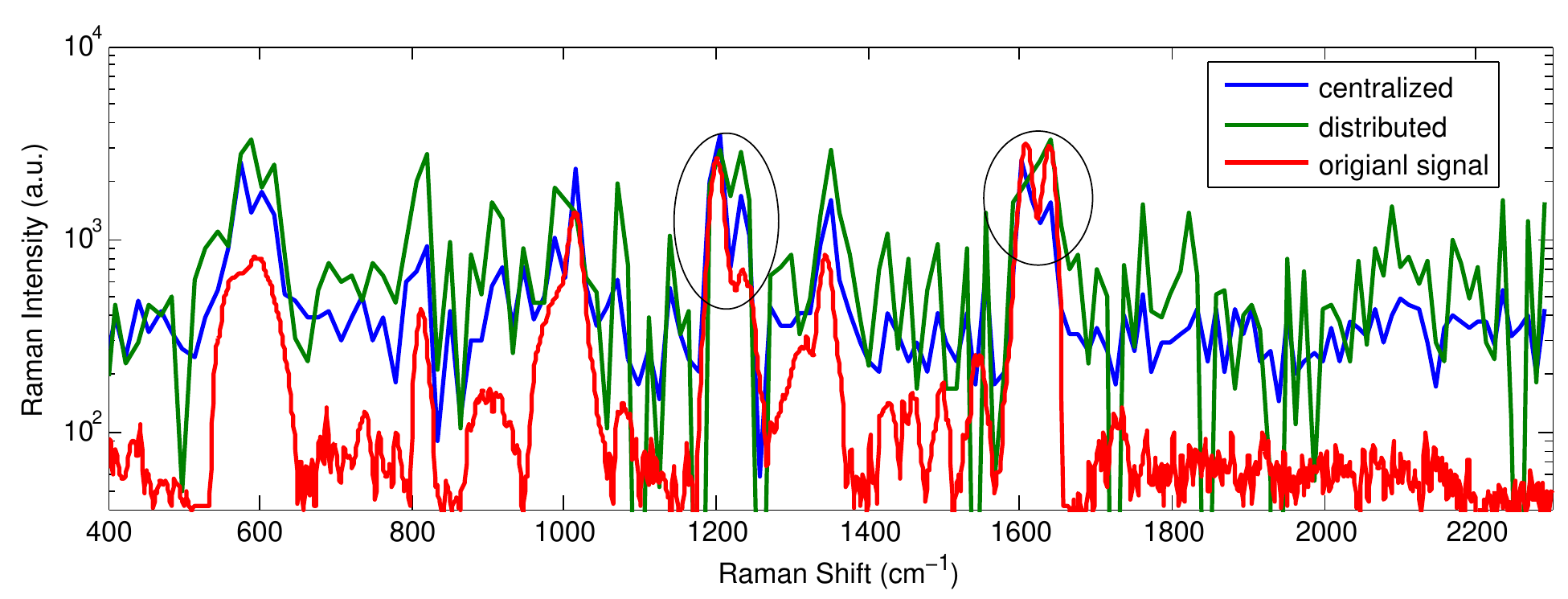}
    \vspace{-5pt}
  \caption{Estimated Raman intensity with low NBP density. Within the left-hand side oval, the estimated signals have two peaks, while the original signal only has one. Within the right-hand side oval, the original signal has two peaks, while the estimated signals have only one.}
  \vspace{-10pt}
  \label{fig:high_threshold}
\end{figure}

As shown in \cite{zhang2015ultrabroadband}, the Raman peaks of 1,2-bits(4-pyridyl)-ethylene molecules are at 1013, 1200, 1342, 1608, and 1636~cm$^{-1}$. As depicted in Fig.~\ref{fig:ideal_large_bw}, by using the centralized sensing architecture the Raman peaks are at 1016, 1205, 1350, 1616, and 1641~cm$^{-1}$ and the MSE is 0.4. By using the distributed sensing architecture the Raman peaks are at 1016, 1205, 1350, 1603, and 1641~cm$^{-1}$ and the MSE is 1.1. The estimated spectrum matches very well with the original spectrum. Moreover, the maximum different of the resonant peaks' Raman shift between the estimated and original signal is 8~cm$^{-1}$. However, if we reduce the transmission power or the NBP density, the accuracy of the estimation results cannot be maintained. For example, in Fig.~\ref{fig:high_threshold} the NBP density is reduced to $1\times 10^{10}$~/s/m$^2$. The MSE of the centralized and distributed sensing results are 1.75 and 2.0, respectively. As we can see in the figure, within the left-hand side oval, the estimated signals have two peaks, while the original signal only has one. Within the right-hand side oval, the original signal has two peaks, while the estimated signals have only one. Due to the low density of NBP, the estimation accuracy is reduced.

 In the following, we investigate the effects of nanosensor number, biofunctional particle density, noise, and transmission power. The outage probability threshold $\tau_t$ is set as 1.5 and 3. If the MSE $e_s$ is smaller than 1.5, we can reconstruct the Raman spectrum accurately. When $1.5\leq e_s \leq 3$, there are some unexpected or missed peaks in the spectrum but the shape of the reconstructed Raman spectrum is still very similar as the original one. When $e_s>3$, the reconstructed Raman spectrum is highly distorted and becomes very different from the original one, which means the results are not acceptable.

\begin{table}
\renewcommand{\arraystretch}{1.3}
\caption{Numerical Parameters}
\label{tab:parameter}
\centering
\begin{tabular}{c|c|c|c|c|c}
    \hline
    Parameter  &  Value & Parameter  &  Value & Parameter  &  Value\\
    \hline
    \hline

    $\lambda_b$  &   $10^6$ /m$^2$ & $u$ & 0.45~m/s & $\lambda_0$ & $2.6\times 10^{10}$/(s $\cdot$ m$^2$ )\\
    \hline

    $S_u $   &   0.003~cm$^2$& $S_l$ & 3$\times$ 10$^{-5}$~ cm$^2$ & $h_c$ & 2.5~mm \\
    \hline

    $r_f$    &   5~mm& $r_b$ & 2.5~mm & $P_t$ & 10~dBm\\
    \hline

    $\alpha$    & $\frac{\pi}{36}$ & $G_s$& 30 dBi & $N_s$ & 30\\
    \hline

    $G_r$    &   30 dBi& $B_w $& 1~THz & $N_f$ & 148\\
    \hline
    $\sigma_m$ & 1 & $\sigma_r$ & 1 &  $\upsilon$ & 1\\
    \hline
\end{tabular}
\end{table}

\subsection{Nanosensor Number}
\begin{figure}[t]
  \centering
    \includegraphics[width=0.6\textwidth]{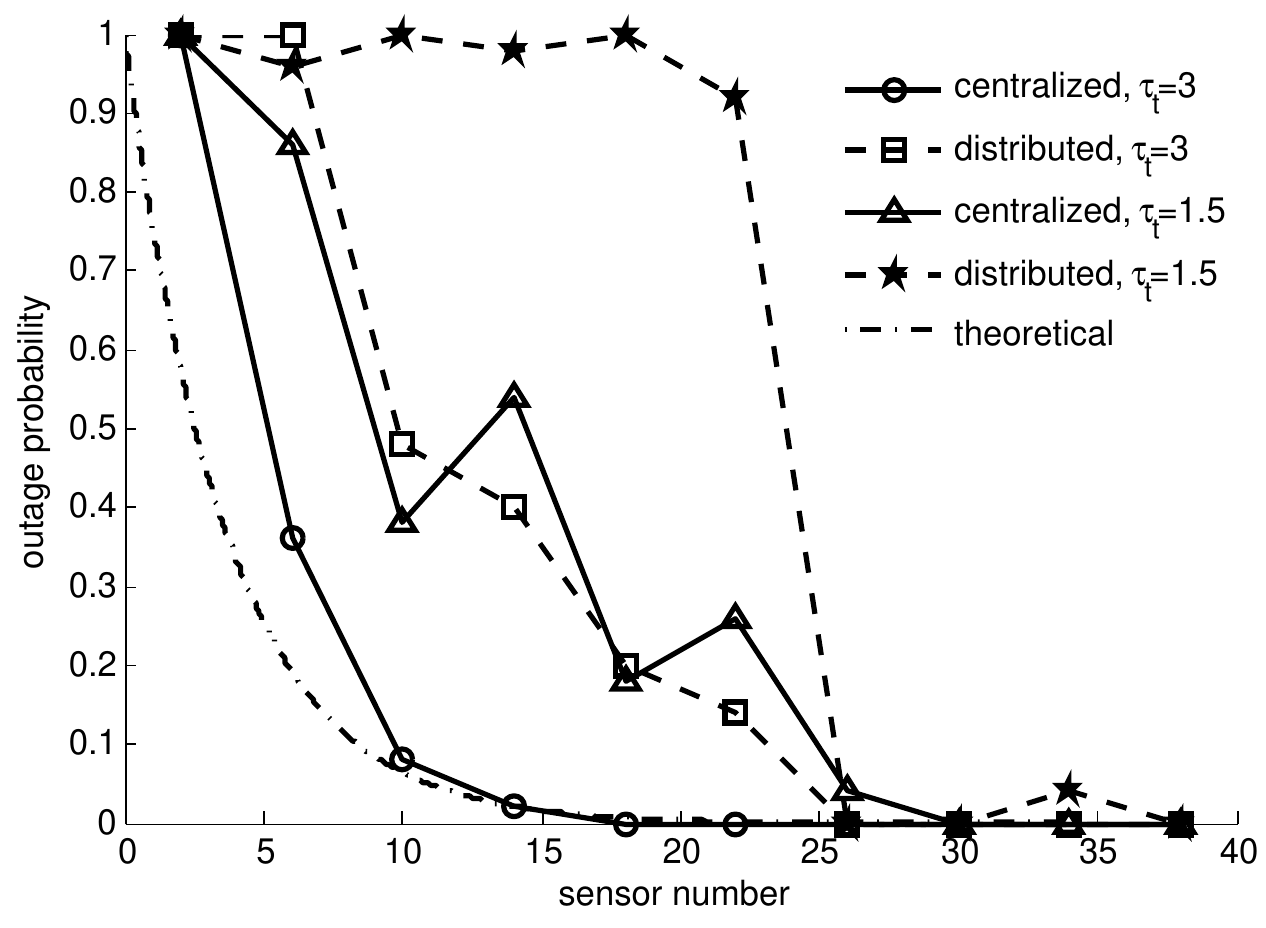}
    \vspace{-5pt}
  \caption{Effect of nanosensor number.}
  \vspace{-10pt}
  \label{fig:sensor_number}
\end{figure}
In \eqref{equ:sensor_number} we derived the minimum nanosensor number based on the blood vessel density. The nanosensor number should satisfy \eqref{equ:sensor_number} to guarantee that there are blood vessels going across the beam cone for all the sub-bands. In Fig.~\ref{fig:sensor_number}, the nanosensor number is varied and the outage probability of the estimation error is evaluated. The threshold $\tau_b$ in \eqref{equ:sensor_number} is set as the same as the outage probability. As we can see in the figure, the theoretical minimum number of nanosensors derived in \eqref{equ:sensor_number} is lower than other estimation outage probability. Hence, it requires fewer nanosensors to satisfy the condition in \eqref{equ:sensor_number}, but more nanosensors are needed to achieve a certain accuracy. Moreover, it is obvious that the centralized sensing architecture requires fewer nanosensors than the distributed sensing architecture. When the nanosensor number is larger than 30, both the centralized and distributed sensing architecture can achieve very high estimation accuracy. Observe that there are some fluctuations on the curves; this is mainly due to the distribution of the nanosensors and some blood vessels in the nano-detectors' beam are far from the detectors which makes the detected power small. As the number of nanosensor increases, this effect decreases.

\subsection{Nano-biofunctional Particle Density}
\begin{figure}[t]
  \centering
    \includegraphics[width=0.6\textwidth]{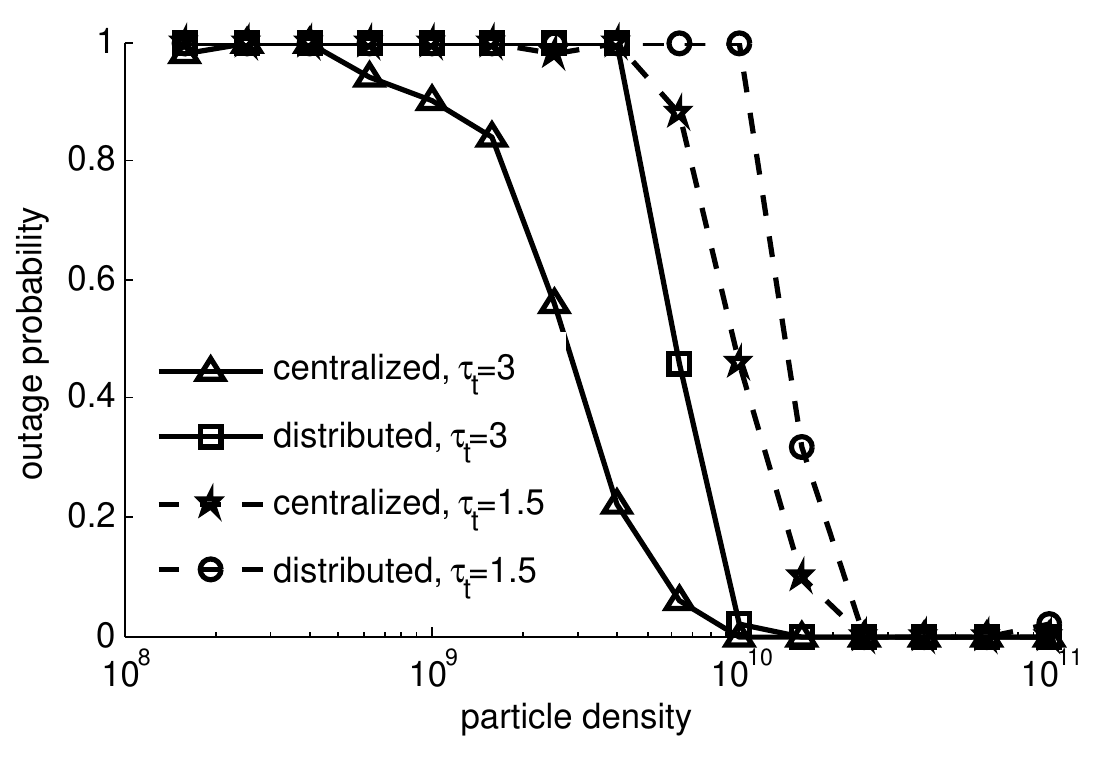}
    \vspace{-5pt}
  \caption{Effect of biofunctional particle density on the outage probability.}
  \vspace{-10pt}
  \label{fig:particle_density}
\end{figure}
The minimum biofunctional particle density is always desired to reduce the side-effects. In Fig.~\ref{fig:particle_density} the density is varied from $10^8$/s/m$^2$ to $10^{11}$/s/m$^2$. Similarly, the centralized sensing architecture still outperforms the distributed sensing architecture, i.e., it requires smaller NBP density. In addition, to achieve near zero outage probability with high estimation accuracy ($\tau_t=1.5$) the required density is $2.6\times 10^{10}$/s/m$^2$ for centralized sensing architecture and distributed sensing architecture which was adopted in the ideal estimation. In addition, we notice that the outage probability of the centralized sensing results decreases gradually with the NBP density increases, while the outage probability of the distributed sensing results drops much faster. They almost require the same NBP density to obtain accurate estimation results. The reason is that when some nanosensors receives highly distorted data, the centralized algorithm can mitigate this effect by averaging the data. However, the distributed sensing architecture first lost a certain accuracy during quantization. Moreover, the weight of the highly distorted data is large in the distributed estimation algorithm since the nanosensors are divided into sub-groups and each nanosensor plays an important role in its sub-group. This effect can be reduced by using more nanosensors.

\subsection{Effect of Noise and Transmission Power}

\begin{figure}[t]
  \centering
    \includegraphics[width=0.6\textwidth]{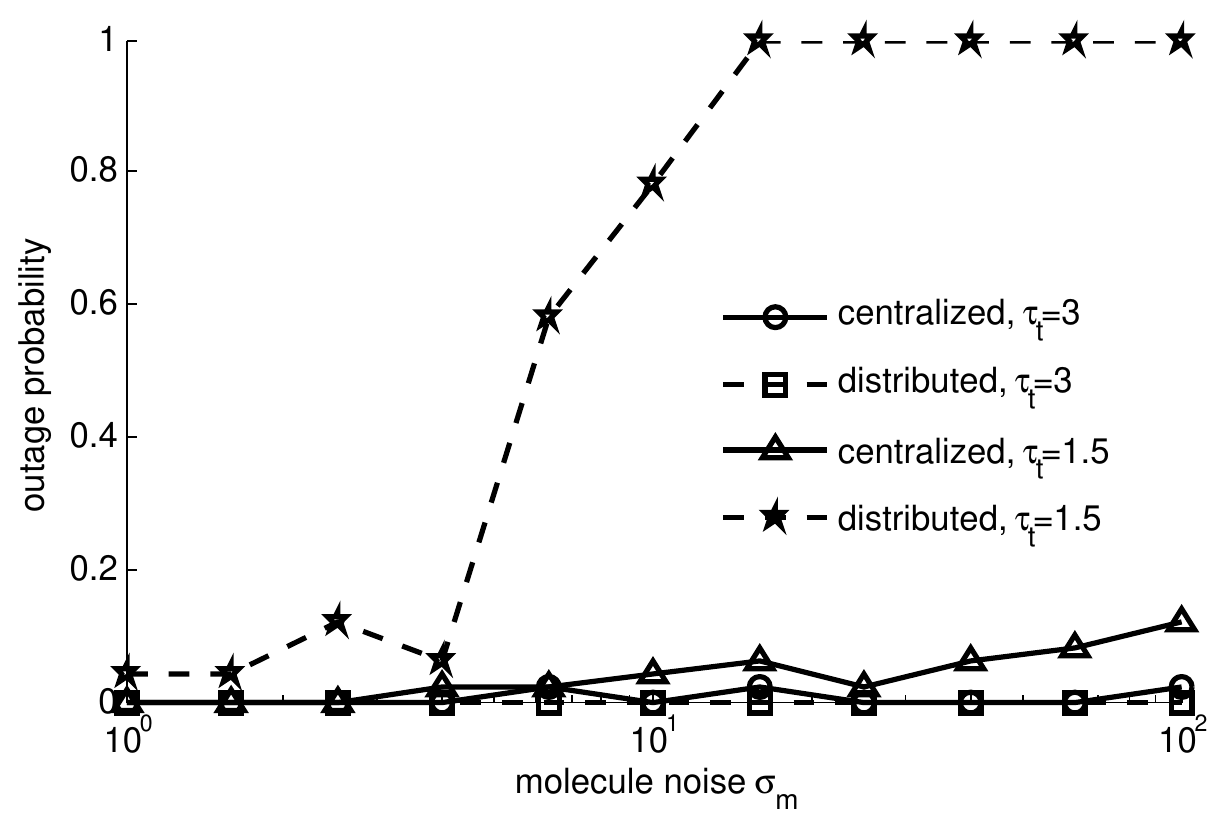}
    \vspace{-5pt}
  \caption{Molecule noise effect on outage probability}
  \vspace{-10pt}
  \label{fig:outage_molecule_noise}
\end{figure}

\begin{figure}[t]
  \centering
    \includegraphics[width=0.6\textwidth]{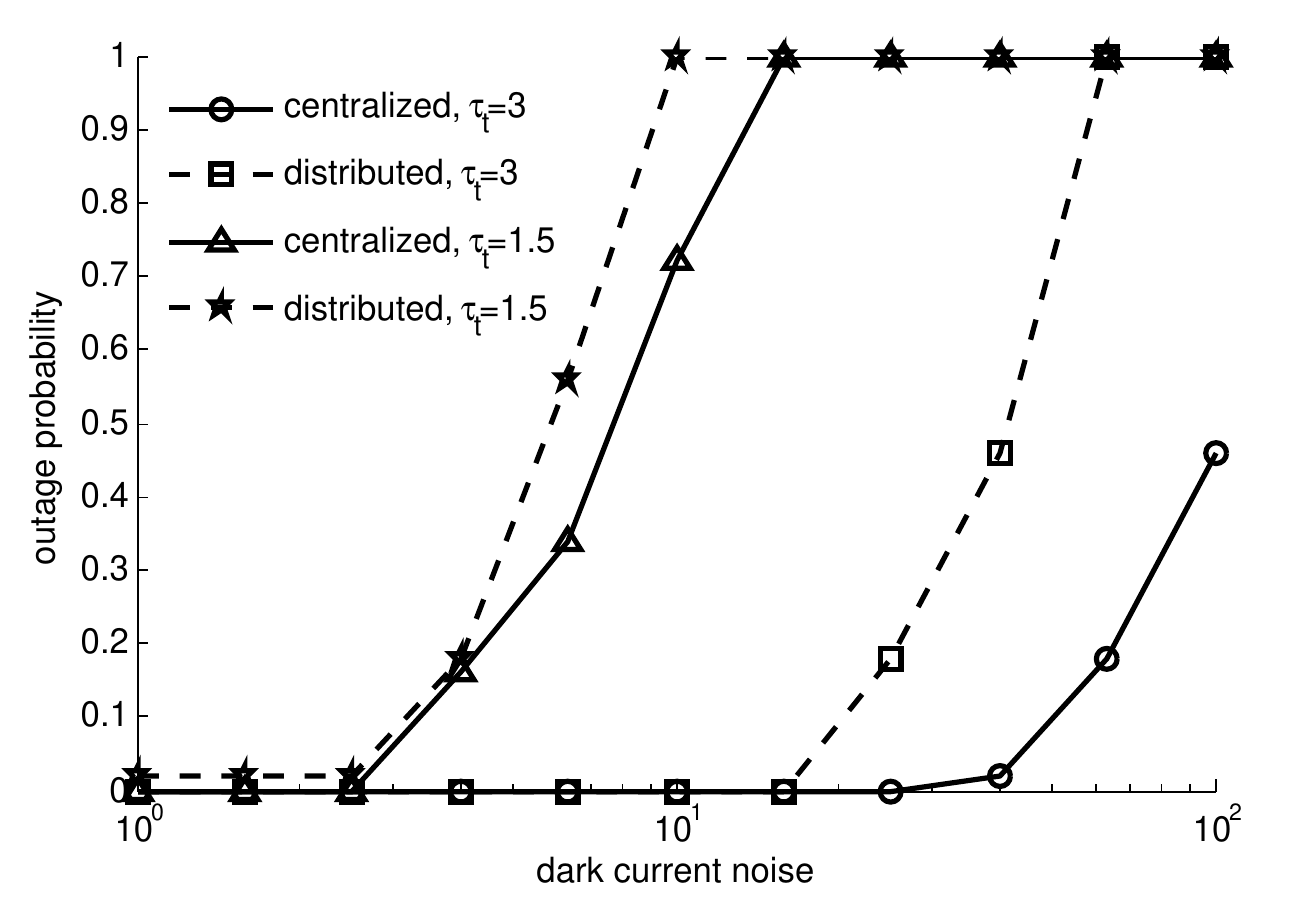}
    \vspace{-5pt}
  \caption{Dark current noise effect on outage probability}
  \vspace{-10pt}
  \label{fig:outage_dark_current_optimal_power}
\end{figure}

\begin{figure}[t]
  \centering
    \includegraphics[width=0.6\textwidth]{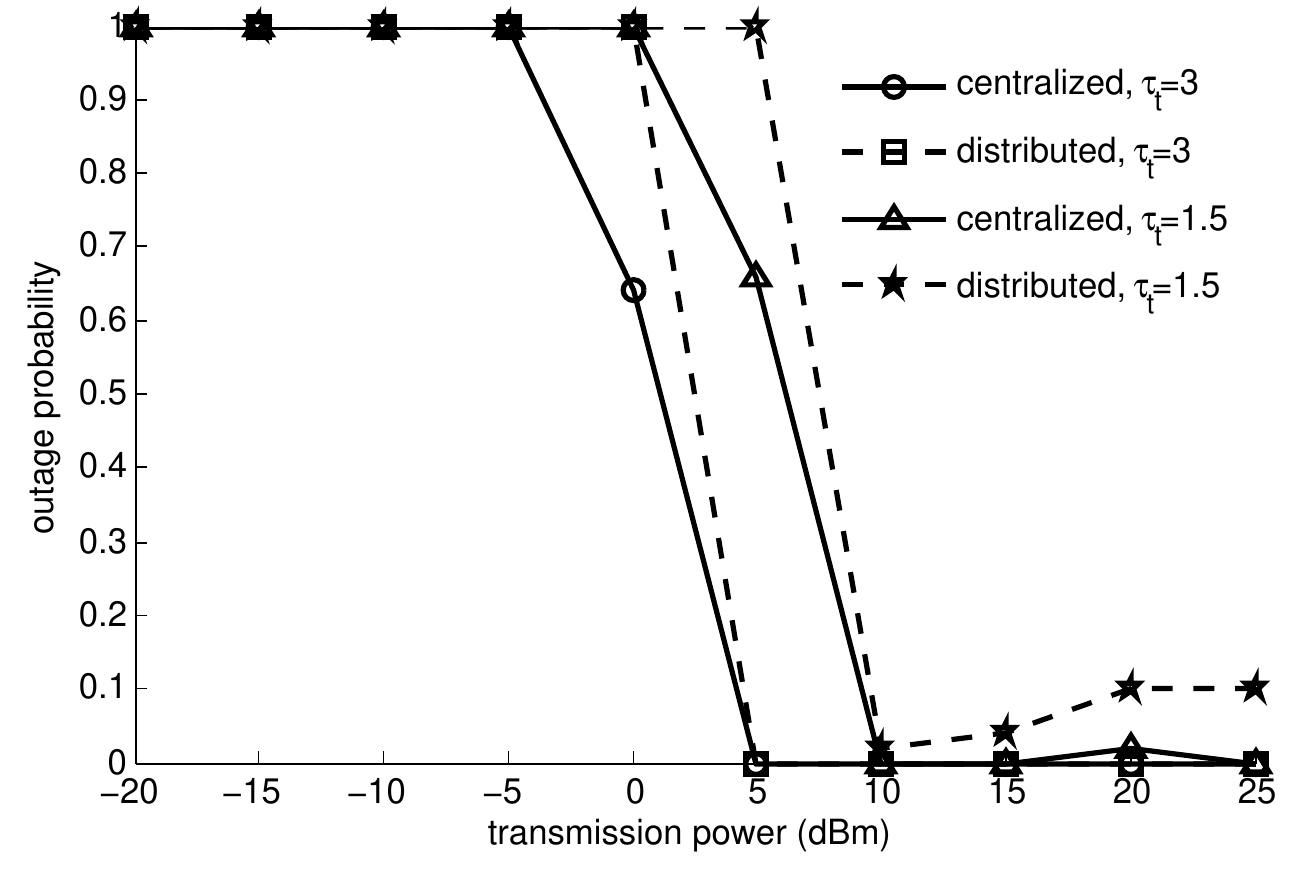}
    \vspace{-5pt}
  \caption{Effect of transmission power.}
  \vspace{-10pt}
  \label{fig:transmission_power}
\end{figure}

The detected signal-to-noise ratio are mainly determined by the noise level and the transmission power. As discussed in preceding sections, the molecule noise and the shot noise (mainly dark current) affect the estimation in different ways. In Fig.~\ref{fig:outage_molecule_noise} the influence of molecule noise is evaluated. As we can see, when $\sigma_m$ is smaller than 4 both the centralized sensing architecture and distributed sensing architecture can achieve very accurate estimation. However, as the noise increases, the distributed sensing architecture becomes inaccurate. Also, when $\sigma_m$ is larger than 25, the centralized sensing architecture with outage threshold 1.5 also increases slowly. Generally, the molecule noise does not have strong influence on the spectrum reconstruction as long as it is not very strong. The reason is that the molecule noise is added together with the scattering coefficient, i.e., $\eta_{f_t,f_j}+\kappa^m$, and the primary feature of the Raman spectrum is resonant peaks. Since $\eta_{f_t,f_j}$ is large at the resonant Raman shift, the noise has negligible effects. As a result, the resonant peaks are not prone to be corrupted by molecule noise.

The effect of shot noise is shown in Fig.~\ref{fig:outage_dark_current_optimal_power}. Different from the molecule noise, shot noise can influence the estimation accuracy dramatically. Here we mainly consider the dark current noise. If the signal power is comparable with the dark current noise, the detected photon number may shift drastically from the accurate value accordingly to \eqref{equ:prob_photon_num}. Moreover, as analyzed in Section~\ref{sec:opt_power}, the dark current noise can create unexpected peaks in the Raman spectrum, which makes the spectrum unrecognizable. On the other hand, we can increase the estimation accuracy by increasing the transmission power. As shown in Fig.~\ref{fig:outage_dark_current_optimal_power}, when the dark current noise is larger than 2.5, the estimation results becomes inaccurate. When it is larger than 25, both centralized and distributed sensing architecture become unacceptable

Next, we evaluate the effect of transmission power. As depicted in Fig.~\ref{fig:transmission_power}, when the transmission power is low the signal is corrupted by the noises in the system and the outage probability is high. For both centralized and distributed sensing architecture, 10~dBm is the minimum amount of required transmission power to achieve high estimation accuracy. We also noted that when the transmission power further increases above 20~dBm, the outage probability of distributed sensing increase slightly. This is because the high received power increases the variance in \eqref{equ:prob_photon_num} which reduces the estimation accuracy. Moreover, although the centralized sensing architecture requires less transmission power, this does not imply that it is more power efficient. Because data communication and quantization also consume power which are not counted here.
\section{Conclusion}
Biosensing using nanotechnology can provide unprecedented accuracy for bio-detection of DNA and proteins, and disease diagnosis and treatment. Although conventional Raman spectroscopy can provide information at nanoscale in intra-body environment, the equipment is bulky and expensive. In this paper, we propose a cooperative Raman spectroscopy using a large number of nanosensors on a smart ring. In this way, the sensing device can be portable and affordable. The nanosensors can jointly and distributively emit and detect optical signals. Meanwhile, the nano-biofunctional particles (NBP) with health information can absorb optical power and then send the information to nano-detectors via Raman scattering. We propose the centralized and distributed sensing architectures to estimate the Raman spectrum. The mathematical models of each component in the sensing system are introduced and the information capacity of the sensing system is derived to optimally allocate power among nano-emitters. The effect of the NBP density and molecule noise are analyzed and the accuracy of the sensing system are evaluated. The results show that the cooperative Raman spectroscopy is able to provide accurate estimation of the Raman spectrum which can be utilized for molecule and chemicals identification. Because of its small profile and low power consumption, we believe the cooperative Raman spectroscopy can find its significant applications in future smart health.

\ifCLASSOPTIONcaptionsoff
 \newpage
\fi
\bibliographystyle{IEEEtran}
\bibliography{ghz}

\begin{thebibliography}{10}
\providecommand{\url}[1]{#1}
\csname url@samestyle\endcsname
\providecommand{\newblock}{\relax}
\providecommand{\bibinfo}[2]{#2}
\providecommand{\BIBentrySTDinterwordspacing}{\spaceskip=0pt\relax}
\providecommand{\BIBentryALTinterwordstretchfactor}{4}
\providecommand{\BIBentryALTinterwordspacing}{\spaceskip=\fontdimen2\font plus
\BIBentryALTinterwordstretchfactor\fontdimen3\font minus
  \fontdimen4\font\relax}
\providecommand{\BIBforeignlanguage}[2]{{%
\expandafter\ifx\csname l@#1\endcsname\relax
\typeout{** WARNING: IEEEtran.bst: No hyphenation pattern has been}%
\typeout{** loaded for the language `#1'. Using the pattern for}%
\typeout{** the default language instead.}%
\else
\language=\csname l@#1\endcsname
\fi
#2}}
\providecommand{\BIBdecl}{\relax}
\BIBdecl

\bibitem{nanoenvironment}
J.~Riu, A.~Maroto, and F.~X. Rius, ``Nanosensors in environmental analysis,''
  \emph{Talanta}, vol.~69, no.~2, pp. 288--301, 2006.

\bibitem{afsharinejad2016performance}
A.~Afsharinejad, A.~Davy, B.~Jennings, and C.~Brennan, ``{Performance analysis
  of plant monitoring nanosensor networks at THz frequencies},'' \emph{IEEE
  Internet of Things Journal}, vol.~3, no.~1, pp. 59--69, 2016.

\bibitem{nguyen2015surface}
H.~H. Nguyen, J.~Park, S.~Kang, and M.~Kim, ``Surface plasmon resonance: a
  versatile technique for biosensor applications,'' \emph{Sensors}, vol.~15,
  no.~5, pp. 10\,481--10\,510, 2015.

\bibitem{eckert2013novel}
M.~A. Eckert, P.~Q. Vu, K.~Zhang, D.~Kang, M.~M. Ali, C.~Xu, and W.~Zhao,
  ``Novel molecular and nanosensors for in vivo sensing,'' \emph{Theranostics},
  vol.~3, no.~8, p. 583, 2013.

\bibitem{ramesh2013towards}
A.~Ramesh, F.~Ren, P.~Berger, P.~Casal, A.~Theiss, S.~Gupta, and S.~Lee,
  ``Towards in vivo biosensors for low-cost protein sensing,''
  \emph{Electronics Letters}, vol.~49, no.~7, pp. 450--451, 2013.

\bibitem{atakan2012body}
B.~Atakan, O.~B. Akan, and S.~Balasubramaniam, ``Body area nanonetworks with
  molecular communications in nanomedicine,'' \emph{IEEE Communications
  Magazine}, vol.~50, no.~1, 2012.

\bibitem{akyildiz2012monaco}
I.~Akyildiz, F.~Fekri, R.~Sivakumar, C.~Forest, and B.~Hammer, ``Monaco:
  fundamentals of molecular nano-communication networks,'' \emph{IEEE Wireless
  Communications}, vol.~19, no.~5, 2012.

\bibitem{vijayarangamuthu2014nanoparticle}
K.~Vijayarangamuthu and S.~Rath, ``Nanoparticle size, oxidation state, and
  sensing response of tin oxide nanopowders using raman spectroscopy,''
  \emph{Journal of Alloys and Compounds}, vol. 610, pp. 706--712, 2014.

\bibitem{saha2012gold}
K.~Saha, S.~S. Agasti, C.~Kim, X.~Li, and V.~M. Rotello, ``Gold nanoparticles
  in chemical and biological sensing,'' \emph{Chemical reviews}, vol. 112,
  no.~5, pp. 2739--2779, 2012.

\bibitem{henry2016surface}
A.-I. Henry, B.~Sharma, M.~F. Cardinal, D.~Kurouski, and R.~P. Van~Duyne,
  ``Surface-enhanced raman spectroscopy biosensing: In vivo diagnostics and
  multimodal imaging,'' \emph{Analytical chemistry}, vol.~88, no.~13, pp.
  6638--6647, 2016.

\bibitem{kneipp1997single}
K.~Kneipp, Y.~Wang, H.~Kneipp, L.~T. Perelman, I.~Itzkan, R.~R. Dasari, and
  M.~S. Feld, ``Single molecule detection using surface-enhanced raman
  scattering (sers),'' \emph{Physical review letters}, vol.~78, no.~9, p. 1667,
  1997.

\bibitem{pantelopoulos2010survey}
A.~Pantelopoulos and N.~G. Bourbakis, ``A survey on wearable sensor-based
  systems for health monitoring and prognosis,'' \emph{IEEE Transactions on
  Systems, Man, and Cybernetics, Part C (Applications and Reviews)}, vol.~40,
  no.~1, pp. 1--12, 2010.

\bibitem{rhee1998ring}
S.~Rhee, B.-H. Yang, K.~Chang, and H.~H. Asada, ``The ring sensor: a new
  ambulatory wearable sensor for twenty-four hour patient monitoring,'' in
  \emph{Engineering in Medicine and Biology Society, 1998. Proceedings of the
  20th Annual International Conference of the IEEE}, vol.~4.\hskip 1em plus
  0.5em minus 0.4em\relax IEEE, 1998, pp. 1906--1909.

\bibitem{feng2014single}
L.~Feng, Z.~J. Wong, R.-M. Ma, Y.~Wang, and X.~Zhang, ``Single-mode laser by
  parity-time symmetry breaking,'' \emph{Science}, vol. 346, no. 6212, pp.
  972--975, 2014.

\bibitem{nafari2017modeling}
M.~Nafari and J.~M. Jornet, ``Modeling and performance analysis of metallic
  plasmonic nano-antennas for wireless optical communication in nanonetworks,''
  \emph{IEEE Access}, 2017.

\bibitem{jacques2013optical}
S.~L. Jacques, ``Optical properties of biological tissues: a review,''
  \emph{Physics in medicine and biology}, vol.~58, no.~11, p. R37, 2013.

\bibitem{prahl1989monte}
S.~A. Prahl, M.~Keijzer, S.~L. Jacques, and A.~J. Welch, ``A monte carlo model
  of light propagation in tissue,'' \emph{Dosimetry of laser radiation in
  medicine and biology}, vol.~5, pp. 102--111, 1989.

\bibitem{wang1995mcml}
L.~Wang, S.~L. Jacques, and L.~Zheng, ``Mcml—monte carlo modeling of light
  transport in multi-layered tissues,'' \emph{Computer methods and programs in
  biomedicine}, vol.~47, no.~2, pp. 131--146, 1995.

\bibitem{lin2011electromagnetic}
J.~C. Lin, \emph{Electromagnetic fields in biological systems}.\hskip 1em plus
  0.5em minus 0.4em\relax CRC press, 2011.

\bibitem{guo2016intra}
H.~Guo, P.~Johari, J.~M. Jornet, and Z.~Sun, ``Intra-body optical channel
  modeling for in vivo wireless nanosensor networks,'' \emph{IEEE transactions
  on nanobioscience}, vol.~15, no.~1, pp. 41--52, 2016.

\bibitem{21285}
A.~D. Wyner, ``Capacity and error exponent for the direct detection photon
  channel. ii,'' \emph{IEEE Transactions on Information Theory}, vol.~34,
  no.~6, pp. 1462--1471, Nov 1988.

\bibitem{shamai1993bounds}
S.~Shamai and A.~Lapidoth, ``Bounds on the capacity of a spectrally constrained
  poisson channel,'' \emph{IEEE Transactions on Information Theory}, vol.~39,
  no.~1, pp. 19--29, 1993.

\bibitem{haas2003capacity}
S.~M. Haas and J.~H. Shapiro, ``Capacity of wireless optical communications,''
  \emph{IEEE Journal on Selected Areas in Communications}, vol.~21, no.~8, pp.
  1346--1357, 2003.

\bibitem{erdmann1978recurrence}
J.~C. Erdmann and R.~I. Gellert, ``Recurrence rate correlation in scattered
  light intensity,'' \emph{JOSA}, vol.~68, no.~6, pp. 787--795, 1978.

\bibitem{martin2000issues}
T.~Martin, E.~Jovanov, and D.~Raskovic, ``Issues in wearable computing for
  medical monitoring applications: a case study of a wearable ecg monitoring
  device,'' in \emph{Wearable Computers, The Fourth International Symposium
  on}.\hskip 1em plus 0.5em minus 0.4em\relax IEEE, 2000, pp. 43--49.

\bibitem{wyner1988capacity}
A.~D. Wyner, ``Capacity and error exponent for the direct detection photon
  channel. i,'' \emph{IEEE Transactions on Information Theory}, vol.~34, no.~6,
  pp. 1449--1461, 1988.

\bibitem{tse2005fundamentals}
D.~Tse and P.~Viswanath, \emph{Fundamentals of wireless communication}.\hskip
  1em plus 0.5em minus 0.4em\relax Cambridge university press, 2005.

\bibitem{ribeiro2006bandwidth}
A.~Ribeiro and G.~B. Giannakis, ``Bandwidth-constrained distributed estimation
  for wireless sensor networks-part ii: Unknown probability density function,''
  \emph{IEEE Transactions on Signal Processing}, vol.~54, no.~7, pp.
  2784--2796, 2006.

\bibitem{zhang2015ultrabroadband}
N.~Zhang, K.~Liu, Z.~Liu, H.~Song, X.~Zeng, D.~Ji, A.~Cheney, S.~Jiang, and
  Q.~Gan, ``Ultrabroadband metasurface for efficient light trapping and
  localization: A universal surface-enhanced raman spectroscopy substrate for
  “all” excitation wavelengths,'' \emph{Advanced Materials Interfaces},
  vol.~2, no.~10, 2015.

\end{thebibliography}
\end{document}